\documentclass[mathleft]{an}
\usepackage{graphicx}
\usepackage{times}
\begin{document}

% The following seven commands are intended for editorial usage and should be ignored by
% the author(s).
\Pagespan{1}{}% Document's page range. 
\Yearpublication{2006}%
\Yearsubmission{2006}%
\Month{11}%   
\Volume{999}%  
\Issue{88}% 
\DOI{This.is/not.aDOI}% 

\title{The Sloan Digital Sky Survey Monitor Telescope Pipeline}

\author{D.L. Tucker\inst{1}\fnmsep\thanks{Corresponding author:
  \email{dtucker@fnal.gov}\newline}
\and S. Kent,\inst{1,2}
\and M.W. Richmond,\inst{3}
\and J. Annis,\inst{1}
\and J.A. Smith,\inst{4,5,6}
\and S.S. Allam,\inst{1,6}
\and C.T. Rodgers,\inst{6}
\and J.L. Stute,\inst{6}
\and J.K. Adelman-McCarthy,\inst{1}
\and J. Brinkmann,\inst{7}
\and M. Doi,\inst{8}
\and D. Finkbeiner,\inst{9,10}\fnmsep\thanks{Hubble Fellow\newline}
\and M. Fukugita,\inst{11}
\and J. Goldston,\inst{10,12}
\and B. Greenway,\inst{1}
\and J.E. Gunn,\inst{10}
\and J.S. Hendry,\inst{1}
\and D.W. Hogg,\inst{13}
\and S.-I. Ichikawa,\inst{14}
\and \v{Z}. Ivezi\'{c},\inst{10, 15}
\and G.R. Knapp,\inst{10}
\and H. Lampeitl,\inst{1,16}
\and B.C. Lee,\inst{1,17}
\and H. Lin,\inst{1}
\and T.A. McKay,\inst{18}
\and A. Merrelli,\inst{19,20}
\and J.A. Munn,\inst{21}
\and E.H. Neilsen, Jr.,\inst{1}
\and H.J. Newberg,\inst{22}
\and G.T. Richards,\inst{10}
\and D.J. Schlegel,\inst{10,17}
\and C. Stoughton,\inst{1}
\and A. Uomoto,\inst{23}
\and B. Yanny\inst{1}
}
\titlerunning{SDSS MTPIPE}
\authorrunning{D.L. Tucker et al.}
\institute{
Fermi National Accelerator Laboratory, P.O. Box 500, Batavia, IL 60510, USA
\and 
Department of Astronomy and Astrophysics, The University of Chicago, 5640 South Ellis Avenue, Chicago, IL 60637, USA
\and 
Physics Department, Rochester Institute of Technology, 85 Lomb Memorial Drive, Rochester, NY 14623-5603, USA
\and 
Deptartment of Physics \& Astronomy, Austin Peay State University, P.O. Box 4608, Clarksville, TN 37044 USA
\and 
Los Alamos National Laboratory, ISR-4, MS D448, Los Alamos, NM 87545-0000, USA
\and 
Department of Physics and Astronomy, University of Wyoming, Laramie, WY 82071, USA
\and 
Apache Point Observatory, P.O. Box 59, Sunspot, NM 88349, USA
\and 
Institute of Astronomy, School of Science, University of Tokyo, Osawa 2-21-1, Mitaka, Tokyo 181-0015, Japan
\and
Harvard-Smithsonian Center for Astrophysics, 60 Garden Street, Cambridge, MA 02138, USA
\and
Princeton University Observatory, Peyton Hall, Princeton, NJ 08544, USA
\and 
Institute for Cosmic Ray Research, University of Tokyo, 5-1-5 Kashiwa, Kashiwa City, Chiba 277-8582, Japan
\and 
University of California at Berkeley, Departments of Physics and Astronomy, 601 Campbell Hall, Berkeley, CA 94720, USA
\and 
Department of Physics, New York University, 4 Washington Place, New York, NY 10003, USA
\and 
National Astronomical Observatory, 2-21-1 Osawa, Mitaka, Tokyo 181-8588, Japan
\and 
Department of Astronomy, University of Washington, Box 351580, Seattle WA 98195-1580 USA
\and
Space Telescope Science Institute, 3700 San Martin Drive, Baltimore, MD 21218, USA
\and
Lawrence Berkeley National Laboratory, 1 Cyclotron Rd, Berkeley CA 94720-8160, USA
\and 
Department of Physics, University of Michigan, 500 East University, Ann Arbor, MI 48109-1120, USA
\and 
Department of Physics, Carnegie Mellon University, 5000 Forbes Avenue, Pittsburgh, PA 15232, USA
\and 
Department of Astronomy, 105-24, California Institute of Technology, 1201 East California Boulevard, Pasadena, CA 91125, USA
\and 
US Naval Observatory, Flagstaff Station, P.O. Box 1149, Flagstaff, AZ 86002, USA
\and 
Department of Physics, Applied Physics, and Astronomy, Rensselaer Polytechnic Institute, 100 Eighth Street, Troy, NY 12180-3590, USA
\and 
Observatories of the Carnegie Institution of Washington, 813 Santa Barbara Street, Pasadena, CA 91101, USA
}

\received{...}
\accepted{...}
\publonline{...}


\abstract{%
The photometric calibration of the Sloan Digital Sky Survey (SDSS) is
a multi-step process which involves data from three different
telescopes: the 1.0-m telescope at the US Naval Observatory (USNO),
Flagstaff Station, Arizona (which was used to establish the SDSS
standard star network); the SDSS 0.5-m Photometric Telescope (PT) at
the Apache Point Observatory (APO), New Mexico (which calculates
nightly extinctions and calibrates secondary patch transfer fields);
and the SDSS 2.5-m telescope at APO (which obtains the imaging data
for the SDSS proper).\\
In this paper, we describe the Monitor Telescope Pipeline, {\tt
MTPIPE}, the software pipeline used in processing the data from the
single-CCD telescopes used in the photometric calibration of the SDSS
(i.e., the USNO 1.0-m and the PT).  We also describe transformation
equations that convert photometry on the USNO-1.0m $u'g'r'i'z'$ system
to photometry the SDSS 2.5m $ugriz$ system and the results of various
validation tests of the {\tt MTPIPE} software.  Further, we discuss
the semi-automated PT factory, which runs {\tt MTPIPE} in the
day-to-day standard SDSS operations at Fermilab.  Finally, we discuss
the use of {\tt MTPIPE} in current SDSS-related projects, including
the Southern $u'g'r'i'z'$ Standard Star project, the $u'g'r'i'z'$ Open
Star Clusters project, and the SDSS extension (SDSS-II).}

\maketitle

\section{Introduction}
\label{sec:intro}

%\clearpage 
\begin{figure*}
\includegraphics[angle=-90,scale=0.7]{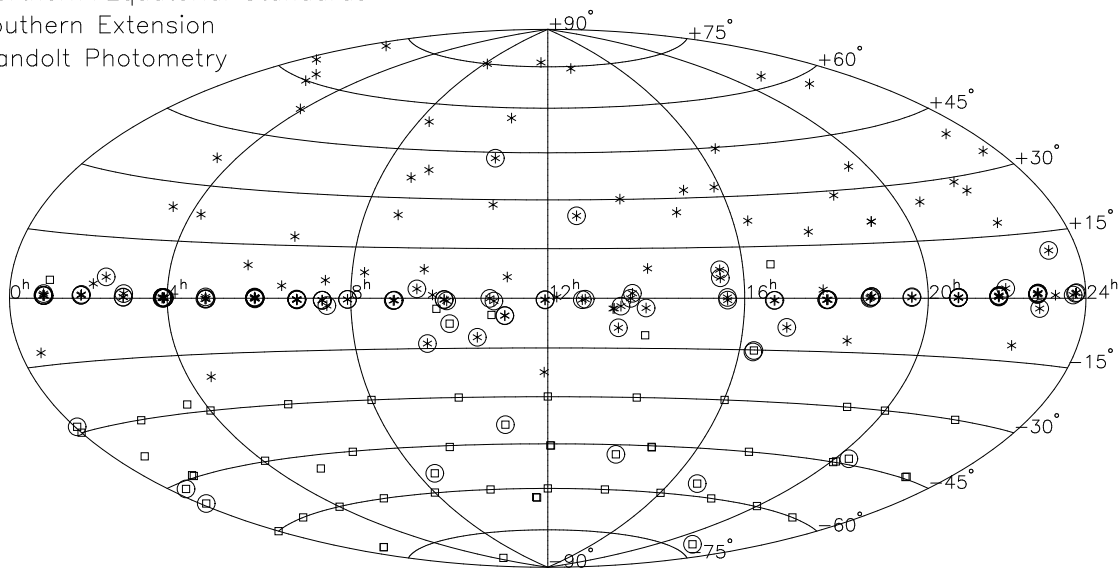}
\caption{Locations of the 158 Smith et al.\ (2002) primary standards (asterisks) 
and the 64 fields of the southern extension to the primary
standard star network (unfilled squares; see
\S~\ref{sec:otherprojects}).  Circled symbols indicate $u'g'r'i'z'$
standard stars or fields for which there is currently Landolt
$UBVR_cI_c$ photometry (Landolt 1973, 1983, 1992) or
for which Landolt is currently obtaining $UBVR_cI_c$ photometry (Landolt, in prep.).
\label{allprifieldlocationsRaDec}} 
\end{figure*}

%\clearpage 
\begin{figure*}
\includegraphics[angle=-90,scale=0.7]{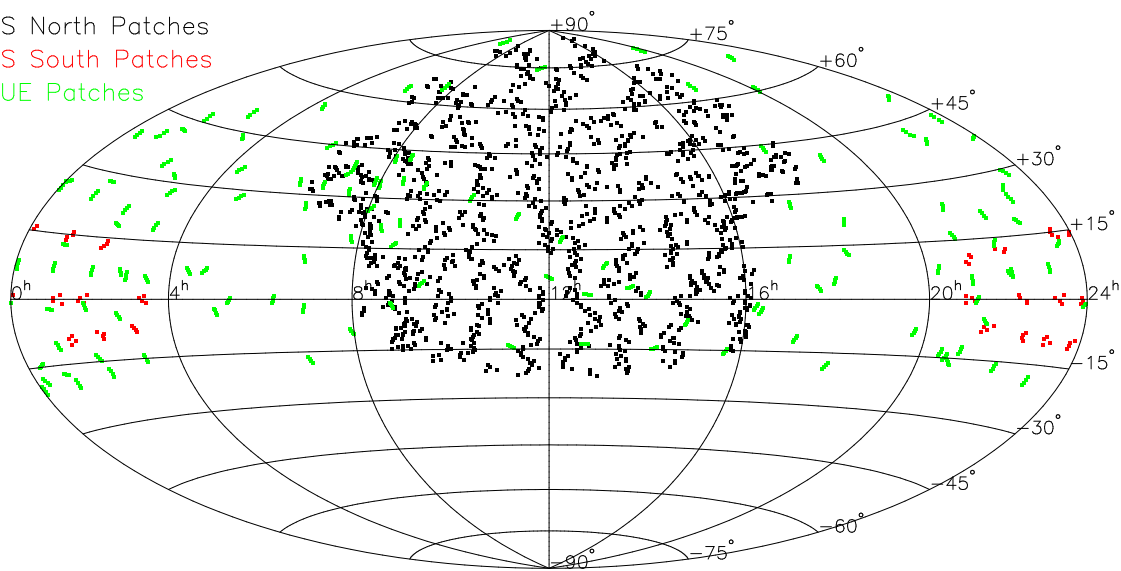}
\caption{ Locations of the SDSS secondary patches in equatorial coordinates 
(Aitoff projection): {\em (black)} Northern SDSS patches, {\em (red)}
Southern SDSS patches, and {\em (green)} SEGUE patches.  (SEGUE is part of the
Sloan extension, or SDSS-II, and is discussed in \S~\ref{sec:otherprojects}.)
\label{secpatchlocationsRaDec}} 
\end{figure*}

The Sloan Digital Sky Survey (SDSS; York et al.\ 2000; Stoughton et
al.\ 2002a; Abazajian et al.\ 2003, 2004, 2005; Adelman-McCarthy et
al.\ 2006) is a modern, CCD-based optical imaging and spectroscopic
survey of the Northern Galactic Cap.  Both imaging and spectroscopy
are performed using a 2.5m f/5 Ritchey-Chr\'etien telescope (Gunn et
al.\ 2006).  The imaging camera (Gunn et al. 1998) contains an imaging
array of thirty 2048$\times$2048 SITe CCDs and scans the sky in drift
scan mode along great circles in five different filters ($ugriz$).  A
complete scan from the western to the eastern borders of the survey
area is called a {\em strip}.  Since there are gaps within the CCD
mosaic, it requires two strips, offset by 93\% of a CCD width, to fill
in a complete rectangular area (which is called {\em stripe}).  The
imaging portion of the SDSS has a depth of $r=22.2$ (95\% detection
repeatablity for point sources; Ivezi\'{c} et al.\ 2004).  Followup
spectrosopy is performed for a variety of targetted galaxies
(Eisenstein et al.\ 2001; Strauss et al. 2002), quasars (Richards et
al. 2002), and stars via a pair of 320-fiber multi-object
spectrographs.

The SDSS imaging covers several thousand square degrees of sky, and
over this region the SDSS photometric calibrations achieve an accuracy
of $\approx$0.02~mag (2\%) rms in $g$, $r$, and $i$ and
$\approx$0.03~mag (3\%) in $u$ and $z$ (Ivezi\'{c} et al.\ 2004;
Adelman-McCarthy et al.\ 2006). The photometric calibration of the
SDSS 2.5m imaging data is a multi-step process which involves the data
from three different telescopes and three different data processing
pipelines.  The telescopes in question are:
\begin{enumerate}

\item the US Naval Observatory (USNO) 1.0-m telescope at Flagstaff 
	Station, Arizona, which was used to set up a network of 158
	primary standard stars for the $u'g'r'i'z'$ photometric system
	in the magnitude range $9 \la r'\la 14$
	(Fig.~\ref{allprifieldlocationsRaDec};
	Fukugita et al.\ 1996; Smith et al.\ 2002);

\item the 0.5-m Photometric Telescope (PT) at Apache Point 
	Observatory (APO), New Mexico, which observes a set of
	$u'g'r'i'z'$ primaries over a range of airmasses to determine
	the photometric solution for the night (zeropoints \&
	extinctions), and calibrates stars down to $r'\approx 18$ in
	transfer fields --- or secondary patches --- that are
	placed throughout the SDSS survey area
	(Fig.~\ref{secpatchlocationsRaDec}); and

\item the SDSS 2.5-m telescope itself (Gunn et al.\ 2006), also located at APO, whose
	imaging camera (Gunn et al.\ 1998) scans over the secondary patches
	(Fig.~\ref{secpatchlocationsRaDec}) during normal course of
	operations.  Note that the imaging camera saturates under
	normal operating conditions at $r \approx 14$, necessitating
	the use of faint stars in the secondary patches (rather than
	observing the primary standards directly).  The secondary
	patches, which are $40\arcmin \times 40\arcmin$ in size, are
	grouped in sets of four, so that each group spans the full
	$2.5^{\circ}$ width of a survey stripe, and each group is
	spaced at roughly $15^{\circ}$ (1~hour) intervals along a
	survey stripe.  [The positions of patches along a stripe have
	been optimized to avoid bright ($V \la 8$) stars; hence
	the slight variability in the spacings between patch groups
	and between patches within patch groups.]  To calibrate the
	2.5m imaging camera photometry, atmospheric extinction
	measurements are taken from PT observations for the night, and
	photometric zeropoints are obtained by matching stars in the
	imaging camera data with those from the secondary patches.
	This method of calibration permits the 2.5m telescope to spend
	its time more efficiently, in that the onus of obtaining
	nightly extinction measurements and of calibrating primary and
	secondary standard stars is shifted to other, smaller
	telescopes.

\end{enumerate}

The three software pipelines are:
\begin{enumerate}

\item the Monitor Telescope Pipeline ({\tt MTPIPE})\footnote{The Monitor
	Telescope Pipeline is named after the original SDSS 0.6m
	calibration telescope at APO, called the Monitor Telescope (or
	MT), which was de-commissioned in Summer 1998 and replaced by
	the PT, which was commissioned in Spring 1999.  Overlapping
	this time frame, the USNO-1.0m telescope was used to set up
	the $u'g'r'i'z'$ standard star network, a task originally
	planned for the old MT.}, which was used to process the
	USNO-1.0m data for setting up the $u'g'r'i'z'$ primary
	standard star network (Smith et al. 2002), and which is currently
	used to process the PT data for determining the atmospheric
	extinctions at APO each night and for calibrating the
	secondary patch fields to the SDSS $ugriz$ system;

\item the Photometric Pipeline ({\tt PHOTO}; Lupton et al.\ 2001; Lupton 2006), 
	which processes the 2.5m imaging camera data, yielding among
	its many outputs accurate astrometry (Pier et al.\ 2003) and
	$ugriz$ instrumental photometry (counts); and

\item the (New) Final Calibrations Pipeline ({\tt NFCALIB}; \S~4.5.3. of
	Stoughton et al.\ 2002a), which matches the stars in the PT
	secondary transfer fields with the 2.5m imaging camera
	photometry, calculates the resulting photometric zeropoints,
	and applies them to the 2.5m instrumental magnitudes.  (To
	deal with 2.5m imaging scans that are observed over a range of
	airmasses, {\tt NFCALIB} also makes use of the first-order
	atmospheric extinction coefficients measured by the PT on the
	same night as a given imaging scan.  Since most 2.5m imaging scans
	only cover a small range in airmass, this is typically a small
	effect.)

\end{enumerate}

%\clearpage 
\begin{figure}
\includegraphics[angle=0,scale=0.35]{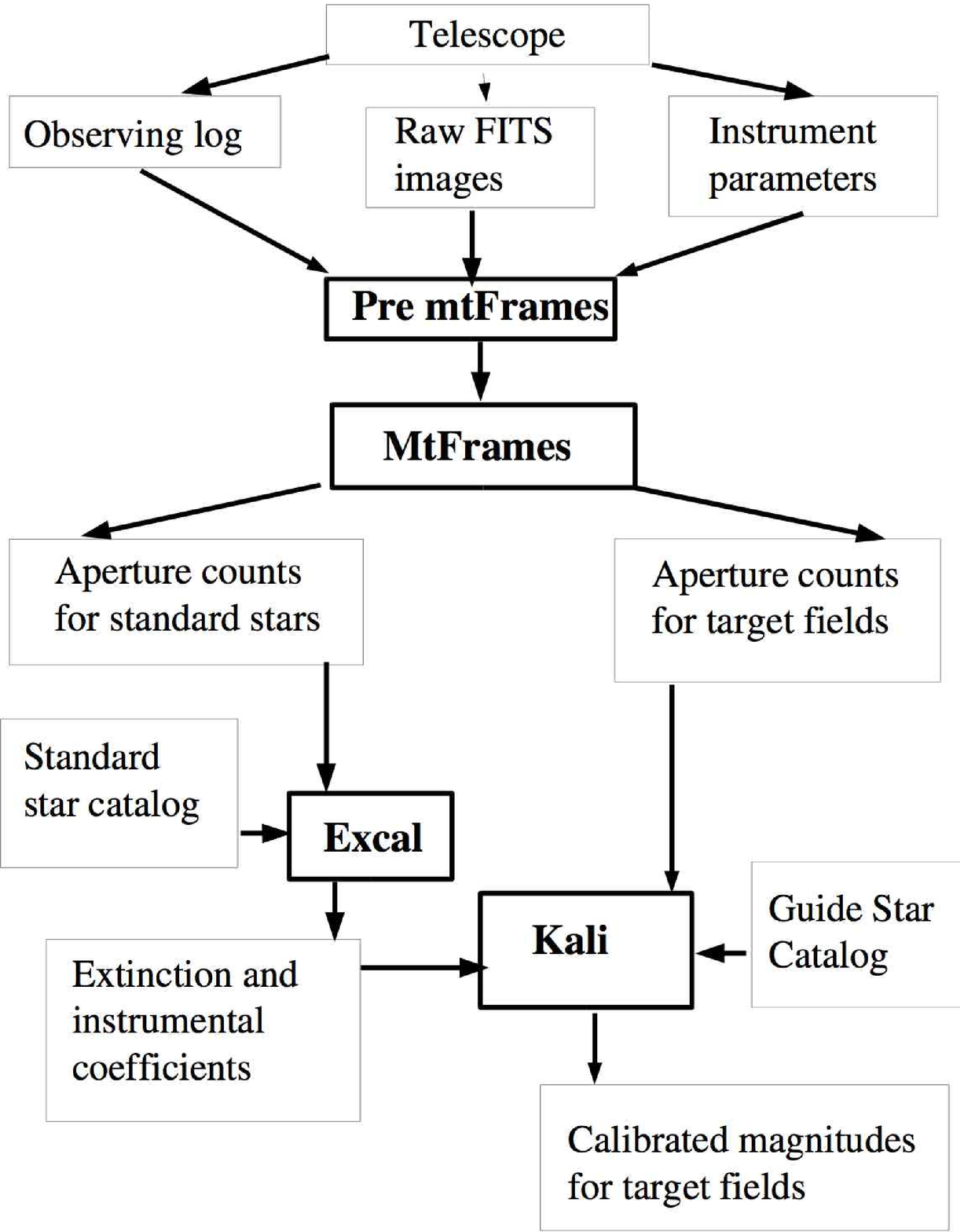}
\caption{A flowchart of {\tt MTPIPE}.
\label{mtpipeFlowChart1}} 
\end{figure}

These pipelines were written with the expressed purpose of meeting the
image processing and calibration needs of the SDSS, and are in no way
meant to supplant the many excellent general-purpose astronomical
image processing and analysis packages like IRAF\footnote{{\tt
iraf.noao.edu}} (Tody 1986, 1993), MIDAS\footnote{{\tt
http://www.eso.org/projects/esomidas/}} (Warmels 1985; Grosb{\o}l
1989), Gypsy\footnote{{\tt http://www.astro.rug.nl/$\sim$gipsy/}} (Shostek
\& Allen 1980; Allen et al. 1992; van der Hulst et al. 1992),
SExtractor (Bertin \& Arnouts 1996), DAOPHOT (Stetson 1987), and
DoPHOT (Schechter et al.\ 1993), to name but a few.  Thus, the SDSS
imaging and photometric calibration pipelines have more in common with
such special-purpose software packages as the ESO Imaging Survey
pipeline (Nonino et al. 1999), the INT Wide Field Camera pipeline
(Irwin \& Lewis 2001), the VLT Survey Telescope pipeline (Grado et
al. 2004), and the Liverpool Telescope Gamma Ray Burst pipeline
(Guidorzi et al. 2006).

The topic of the current paper is {\tt MTPIPE}.  In the following
sections, we will discuss in turn {\tt MTPIPE}'s component packages
(\S~\ref{sec:packages}), validation tests of the {\tt MTPIPE} software
(\S~\ref{sec:tests}), the {\tt MTPIPE}-based semi-automated PT data
processing factory (\S~\ref{sec:ptfactory}), and other SDSS-related
projects using {\tt MTPIPE} (\S~\ref{sec:otherprojects}); in
\S~\ref{sec:conclusions} we comment on future plans for {\tt MTPIPE}.

In closing, we note that, although the SDSS magnitudes and colors are
shown without primes ($ugriz$), the magnitudes of the primary standard
stars are shown with primes ($u'g'r'i'z'$).  This is no accident.  The
primed system is defined in the natural system of the USNO-1.0m, its
CCD, and its set of SDSS filters.  The SDSS magnitudes, however, are
defined in the natural system of the SDSS 2.5m imaging camera and its
SDSS filters.  These two systems are very similar and the coefficients
of the transformation equations are quite small.  We will describe the
differences in these two systems and provide transformation equations
in \S~\ref{sec:packages}.

\section{The {\tt MTPIPE} packages}
\label{sec:packages}

{\tt MTPIPE} is a suite of code written in a combination of the Tcl
and C programming languages and based upon the SDSS {\tt
DERVISH}$+${\tt ASTROTOOLS} (Stoughton 1995; Sergey et al.\ 1996)
software environment.  The current version of {\tt MTPIPE} is {\tt
v8.3}, although changes since {\tt v8.0} have been mostly cosmetic and
have generally dealt with the smooth running of the PT Factory
(\S~\ref{sec:ptfactory}).  {\tt MTPIPE} includes four main packages
used in normal reductions: {\tt preMtFrames}, {\tt mtFrames}, {\tt
excal}, and {\tt kali}, and they are run in that order
(Fig.~\ref{mtpipeFlowChart1}).

\subsection{\tt preMtFrames}

%\clearpage 
\begin{figure*}
\includegraphics[angle=0,scale=0.75]{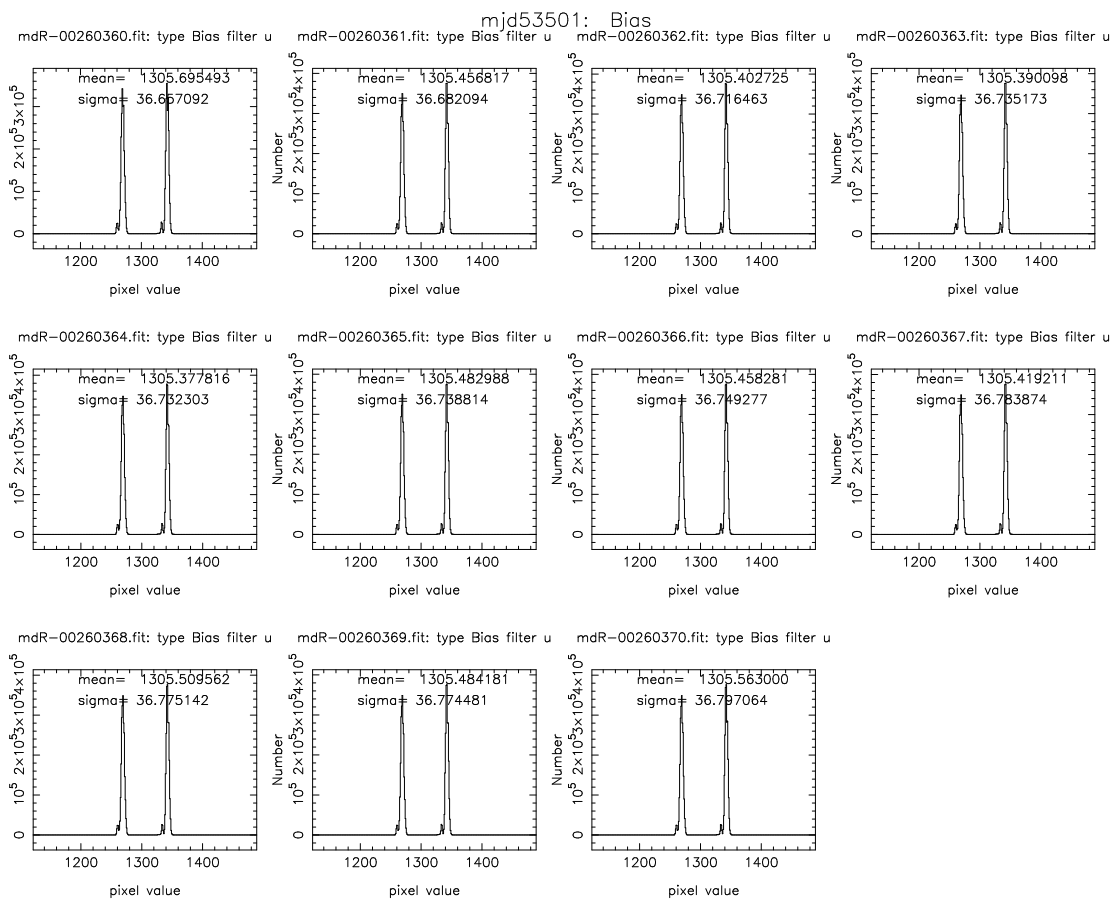}
\caption{Histograms of pixel values for each of the raw PT bias frames taken on MJD 53501.
Note that the PT's CCD has 2-amplifier readout electronics; hence, the bimodality
of the distributions.  
\label{hgBiasFrames53501}} 
\end{figure*}

The first package, {\tt preMtFrames}, is basically a ``pre-burner'':
it creates the directory structure for the reduction of a night's
data, including various parameter files needed as input for the other
three packages, and it runs quality assurance tests on the raw data.
Furthermore, it verifies information in the electronically generated
observing log (which is based upon the information within the FITS
image headers), identifies the type of each image (e.g., bias frame,
dome flat, twilight flat, primary standard field, survey secondary
patch field, or specially targetted manual fields), matches the frame
identifier to a list of approved standard fields, verifies that a full
set of frames ($u'g'r'i'z'$) are present for each target sequence, and
creates histograms for each of the bias and flat field frames.

As an example, plotted in Figure~\ref{hgBiasFrames53501} are
histograms showing the distribution of pixel values for each of the
bias frames observed by the PT on MJD 53501.\footnote{MJD is the
Modified Julian Date, defined by the relation MJD $\equiv$ JD $-$
2,400,000.5, where JD is the Julian Date.}  Note that the PT CCD uses
two-amplifier readout, as evidenced by the strongly bimodal
distribution of pixel values for these bias frames.

Note that {\tt MTPIPE} does not use the values of image FITS header
keywords directly, but makes use of an ASCII observing log (called the
mdReport file) containing the relevant information (e.g., RA, DEC,
time of observation, exposure time, filter, ...) for each image.  This
information is usually based upon image FITS header keyword values,
but can also be user-generated.  The use of the mdReport file
simplifies the process of editing this information by humans when the
need arises.

\subsection{\tt mtFrames}

%\clearpage 
\begin{figure*}
\begin{picture}(400,550)
\put (  0,   0){\includegraphics[angle=0,scale=0.75]{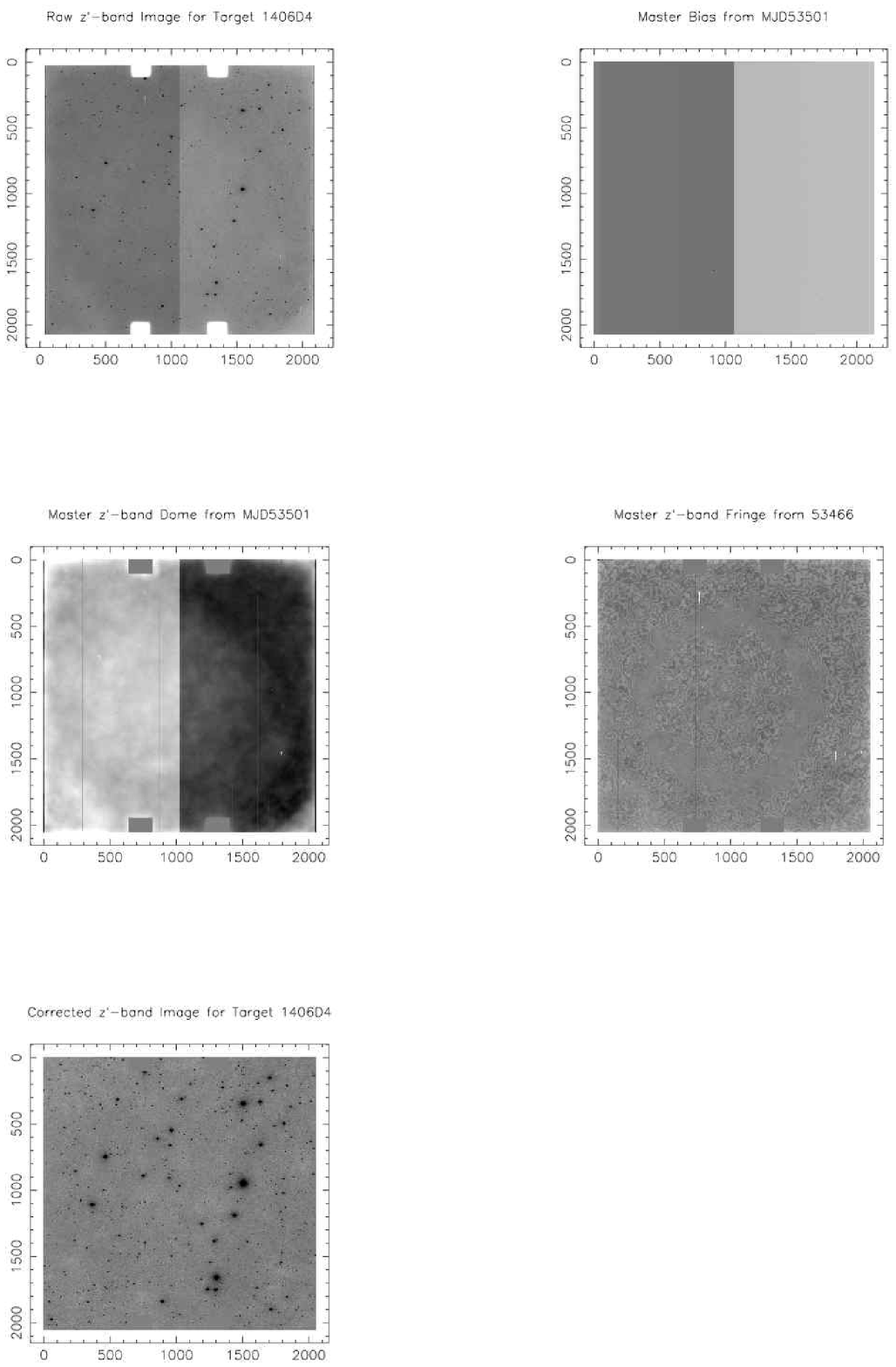}}
\put ( 10, 530) { (a)}
\put (220, 530) { (b)}
\put ( 10, 340) { (c)}
\put (220, 340) { (d)}
\put ( 10, 150) { (e)}
\end{picture}
%\vspace{-1.5cm}
\vspace{-1.0cm}
\caption{The image processing steps, using PT data from MJD 53501 as an example:  
(a) the raw $z'$-band image of the secondary patch 1406D4, (b) the
master bias frame from MJD 53501, (c) the master $z'$-band dome flat from
MJD 53501, (d) master $z'$-band fringe frame from MJD 53466 (used for
processing the PT data from MJD 53501), (e) the bias-subtracted,
trimmed, flat-fielded, and fringe-corrected $z'$-band image of the
secondary patch 1406D4.  The (linear) stretch used in these images is
from 90\% to 110\% of the given image's median pixel value.
\label{processedImages53501}} 
\end{figure*}
%\clearpage 
\begin{figure*}
\begin{picture}(400,220)
%\put (  30, 220){\includegraphics[angle=-90,scale=0.60]{fg06.eps}}
\put (   0, 250){\includegraphics[angle=-90,scale=0.60]{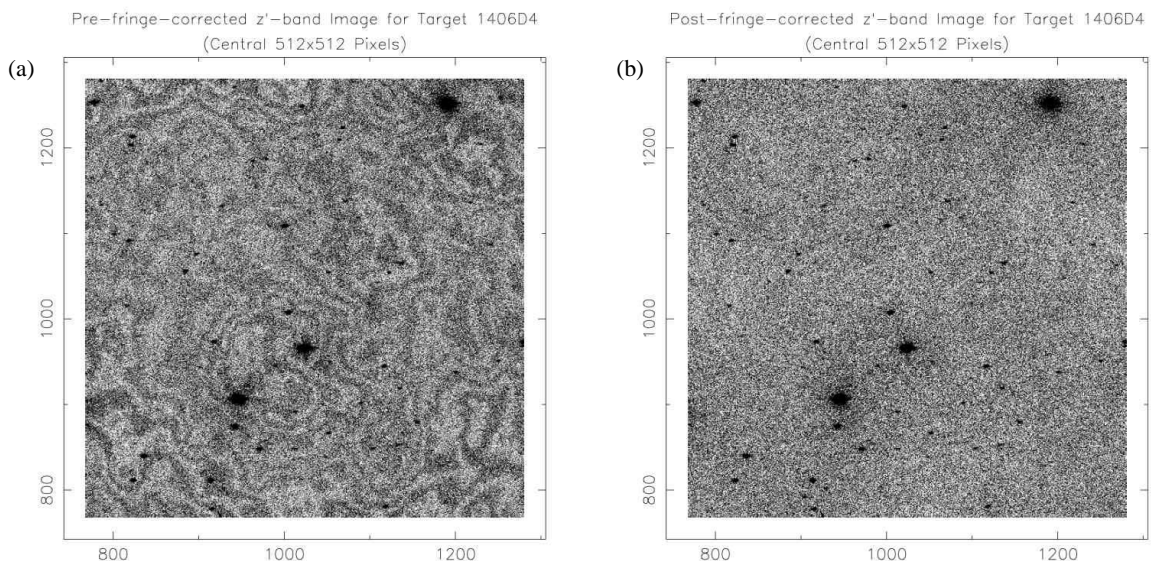}}
\put (  10, 150) { (a)}
\put ( 240, 150) { (b)}
\end{picture}
\vspace{1.25cm}
\caption{An example of de-fringing a PT $z'$-band image:  
(a) the central $512\times512$ pixels of the bias-subtracted, trimmed,
flat-fielded, but not fringe-corrected $z'$-band image of the field
featured in Figure~\ref{processedImages53501}; (b) the same as (a),
but after fringe-correction.  The (linear) stretch in these two images
is the same (from 95\% to 105\% of the median pixel value of (a)).
The typical trough-to-peak amplitudes of the fringes in (a) is
$\approx$2\%.
\label{deFringing}} 
\end{figure*}
%\clearpage 
\begin{figure*}
\begin{picture}(400,600)
\put ( 50, 625){\includegraphics[angle=-90,scale=0.55]{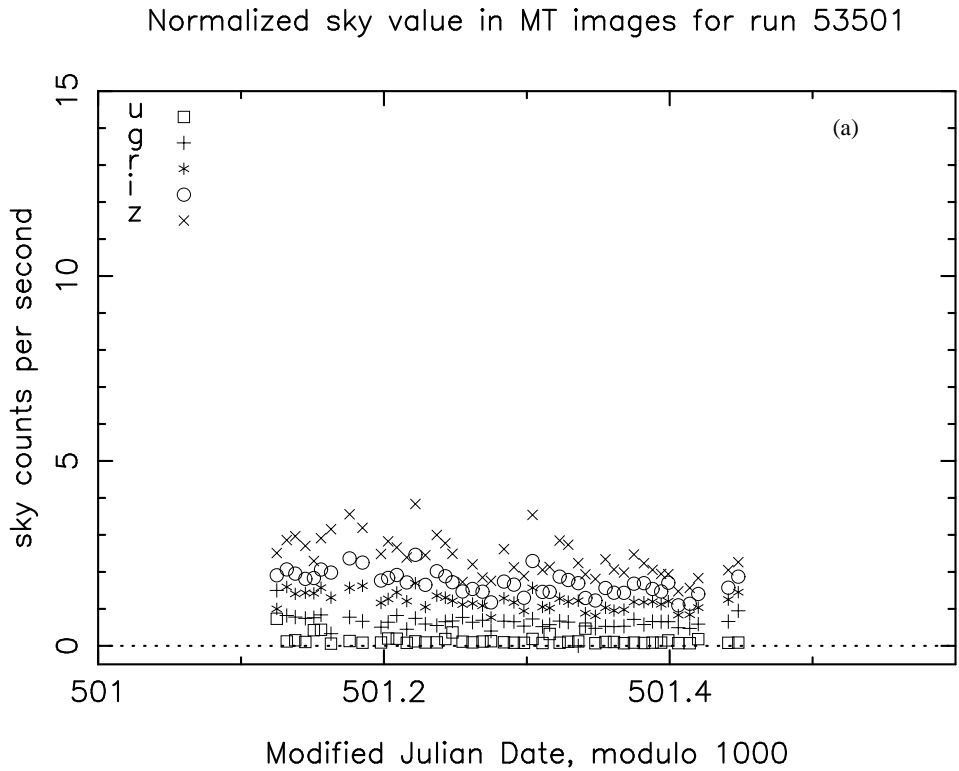}}
\put ( 50, 300){\includegraphics[angle=-90,scale=0.55]{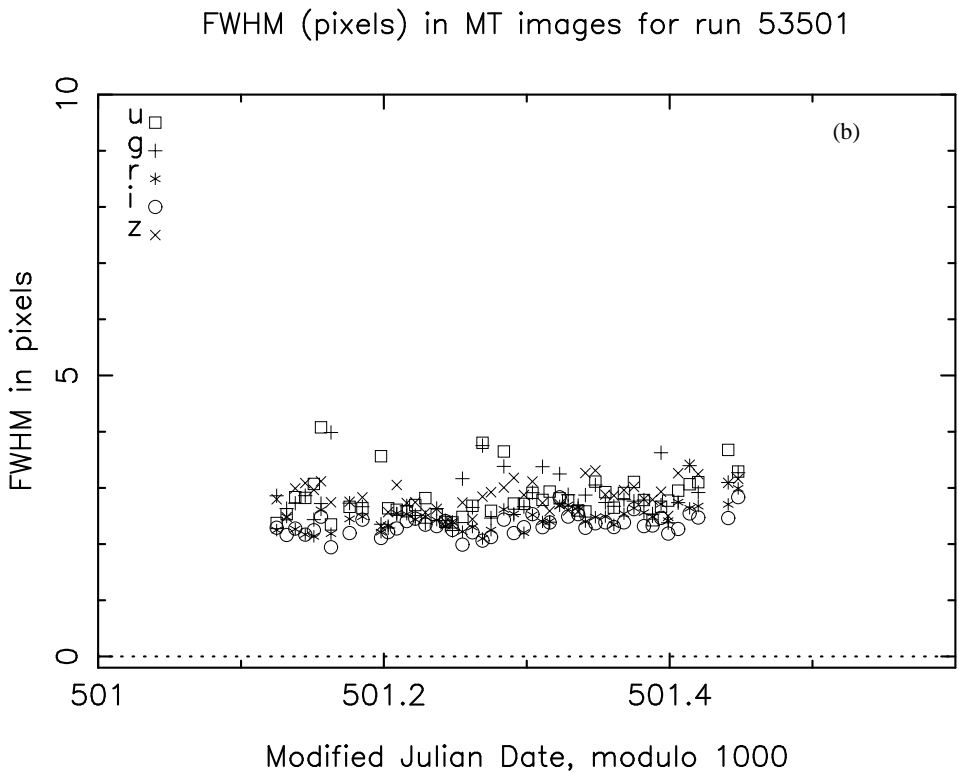}}
\put (360, 575) { (a)}
\put (360, 250) { (b)}
\end{picture}
%\begin{picture}(400,200)
%\put (  0, 180){\includegraphics[angle=-90,scale=0.33]{fg07a.ps}}
%\put (250, 180){\includegraphics[angle=-90,scale=0.33]{fg07b.ps}}
%\put (190, 150) { (a)}
%\put (440, 150) { (b)}
%\end{picture}
\caption{A sampling of the quality assurance plots output by {\tt mtFrames} 
for the PT data taken on the night of MJD 53501: (a) Sky counts (in
ADU/pixel) vs.\ time in each of the five passbands; (b) Mean image
FWHM (in pixels) vs.\ time in each of the five passbands.  (Note that
the scale of the PT's CCD is $1.15\arcsec$/pixel.)
\label{mtFramesQA}} 
\end{figure*}

The next package, {\tt mtFrames}, is the primary image processing
portion of the software, and is capable of processing data from CCDs
with 1-, 2-, or 4-amplifier readout electronics.

The first part of {\tt mtFrames} creates master bias frames, flat
field frames, and fringe frames.  The master biases are created by
median filtering a set raw bias frames.  During general operations
with the PT, typically ten raw bias frames are obtained per night.

Once the master bias is created, master dome flats for each filter are
made; this is done by subtracting the master bias from each raw dome
flat frame, removing any residual median offset from the overscan
regions, trimming off the overscan region, applying any linearity
corrections, and median filtering the thus-processed individual dome
flats for a given filter.  For the PT, typically five raw dome flats
are obtained for each of the five filters ($u'g'r'i'z'$) every afternoon
of scheduled observing.  The same processing steps are employed to
create master twilight flats from the raw twilight flat frames.  For
the PT, raw twilights are obtained mostly in $u'$, since we find that
dome flats sufficiently flatten the PT $g'r'i'z'$ frames, whereas only
twilight flats are capable of adequately flattening the PT $u'$
frames. Typically 5-10 raw twilights (mostly in $u'$) are obtained
each evening twilight in which the skies are reasonably clear.

Master fringe frames are needed to remove the additive effects of
internal CCD illumination patterns seen in the $i'$ and $z'$ filters.
To create master $i'$ and $z'$ fringe frames, {\tt mtFrames} seeks out
all the secondary patches observed during a night.  These secondary
patches are useful for creating master fringe frames because they are
relatively deep (i.e., have relatively long exposure times), and
because they are of relatively uncrowded star fields containing few if
any bright (highly saturated) stars, and because a given secondary
patch is typically not repeat-targetted during a given night.  If
there are a sufficient number of secondary patches observed on that
night (for the PT, the default minimum is 7), master fringe frames
will be created for the $i'$ and $z'$ filters.  The $i'$ and $z'$
frames for the secondary patches are fully processed up through
flatfielding --- they are bias-subtracted using the master bias, their
residual overscan offset is subtracted, their overscan is trimmed,
they are linearity corrected, and they are flatfielded with the
appropriate master flat frame.  After this processing, the secondary
patch frames for a given filter are scaled to a default median
background sky and median filtered.  This final median-filtered image
is the fringe frame for that filter.

Note that, in the above description of creating master biases, flats,
and fringes, we only make mention of data obtained over a single
afternoon+night.  Why not use the data obtained over a full week or
month or more to beat down statistics?  Actually, it is possible to do
this with {\tt mtFrames}, if one collects all the data from a
multiple-night run into a single directory and manipulates the ASCII
observing log file appropriately; this process is currently not
automated, and hence requires a certain level of human interaction.
With general PT operations for the SDSS, however, the guiding
principle is that of considering each night as a self-contained
experiment.  Often, obtaining photometric calibration data for the
SDSS is time critical, and it is ill-advised to wait until the end of
a dark run to obtain (hopefully) all the necessary bias, flat, and
fringe frames.  That said, those nights when it is impossible to
obtain a sufficient set of biases, flats, or (which is more often the
case) secondary patches for fringe frames, the necessary master frames
are copied from another night's processed data.  This is permissible
since the PT's biases and flats are stable over the course of a
typical 3-week observing run --- significant discontinuous changes in
these calibration frames are, however, often noticeable from one
observing run to the next --- and the PT $i'$ and $z'$ band fringe
patterns --- being an artifact of physical variations in the thickness
of the CCD itself --- are stable over even longer periods.

The second part of {\tt mtFrames} applies these master biases, flats,
and fringe frames to the target frames --- i.e., to the frames of
fields containing primary standard stars, of secondary patch fields,
and of any manually targetted fields of interest.  The standard
processing steps apply: for a target frame in a given filter, the
master bias is subtracted, any residual median offset from the
overscan regions is subtracted, the frame is trimmed of the overscan
regions, a linearity correction is applied, and the appropriate master
flat frame is used; if the target frame is in the $i'$ or $z'$ filter,
a fringe frame is subtracted and the median sky value of pre-defringed
frame is added back to all the pixels.

Once a target frame is fully processed, objects are detected as peaks
$n \sigma$ above the sky background, where $\sigma$ is the rms scatter
in the sky background, and aperture photometry is performed (for the
general PT operations, $n=10$).  For the aperture photometry, one of
two apertures may be used.  The larger aperture is used for fields
containing standard stars and for bright stars in secondary patches
and manual target-of-opportunity fields.  The smaller aperture is used
for faint stars in the secondary patches and manual fields; an
aperture correction is applied to convert the small aperture counts
into a measure of the large aperture counts for these faint stars.
The actual sizes of these apertures are parameters that can be
adjusted.  For example, when setting up the original standard star
network (Smith et al.\ 2002), the primary standard stars were
extracted using a 24-arcsec diameter aperture.  This size was selected
to avoid problems associated with defocussing the brightest stars,
required for some of the USNO observations.

Typically a given field is observed in multiple filters before moving
on to the next field, yielding a sequence of target frames for that
field, one for each filter.  In general PT operations, each target
field is observed in all five filters ($u'g'r'i'z'$) before moving on
to the next target field.  Once {\tt mtFrames} has performed aperture
photometry on all the frames for a given $u'g'r'i'z'$ sequence,
objects detected in each filter are matched (by default, they are
matched to the detected objects in the $r'$ filter) and the merged
object list is written to disk as a FITS binary table file.  For the
PT, which has a known radial distortion, a distortion correction is
applied at this stage to the aperture photometry.

As an example of the processing steps in {\tt mtFrames}, we continue
our examination of the PT data from MJD 53501.  In
Figure~\ref{processedImages53501}, we see a raw $z'$ image for one of
the secondary patches observed that night
(Fig.~\ref{processedImages53501}a), the master bias for that night
(Fig.~\ref{processedImages53501}b), the master $z'$-band dome flat for
that night, (Fig.~\ref{processedImages53501}c), the master $z'$-band
fringe frame for that night (actually taken from MJD 53466, due to an
insufficient number of secondary patches observed on MJD 53501;
Fig.~\ref{processedImages53501}d), and the final, bias-subtracted,
flat-fielded, de-fringed $z'$ corrected frame for that secondary patch
(Fig.~\ref{processedImages53501}e).  In Figure~\ref{deFringing}, we
zoom in on the central quarter of the pre- and post-fringe-corrected
frame to show the effects of de-fringing.

Finally, {\tt mtFrames} outputs quality assurance plots for the night
processed.  Two examples are shown in Figure~\ref{mtFramesQA}:
Figure~\ref{mtFramesQA}a and b show, respectively, the sky brightness
(in ADU counts per pixel) vs.\ time and the mean FWHM per image vs.\
time for PT data obtained on the night of MJD 53501.

\subsection{\tt excal}

%\clearpage 
\begin{figure*}
\begin{picture}(400,600)
\put ( 50, 625){\includegraphics[angle=-90,scale=0.55]{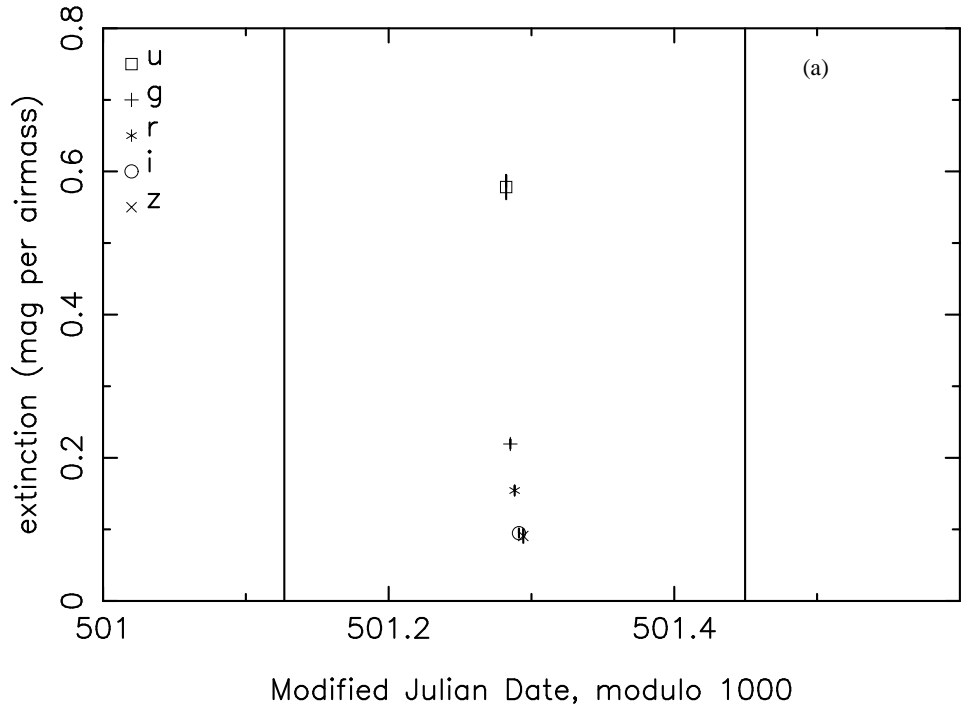}}
\put ( 50, 300){\includegraphics[angle=-90,scale=0.55]{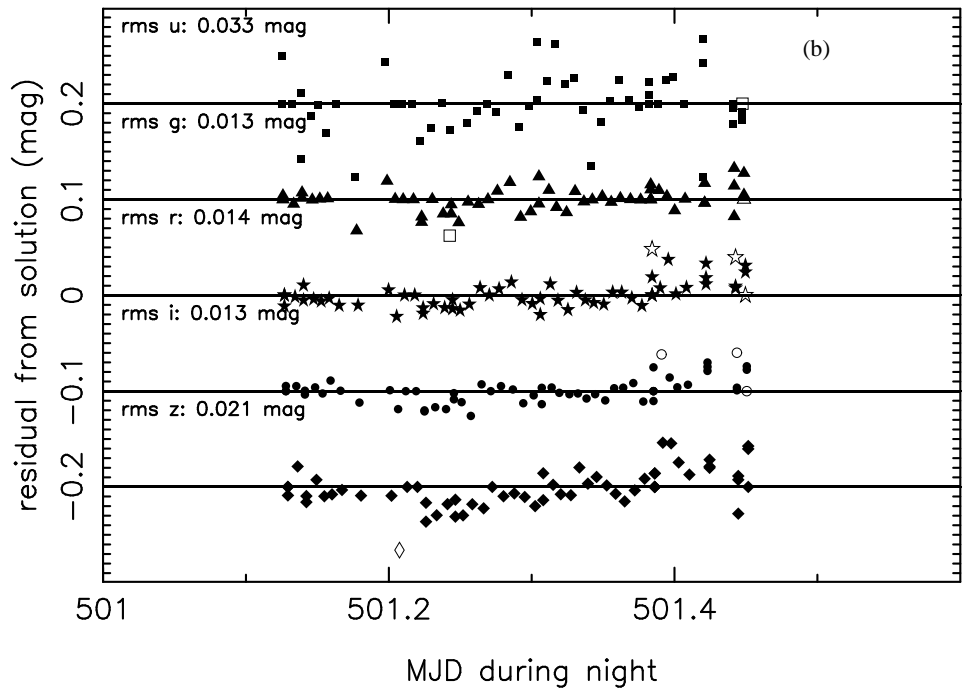}}
\put (360, 575) { (a)}
\put (360, 250) { (b)}
\end{picture}
%\begin{picture}(400,400)
%\put (  0, 400){\includegraphics[angle=-90,scale=0.33]{fg08a.ps}}
%\put (250, 400){\includegraphics[angle=-90,scale=0.33]{fg08b.ps}}
%\put (  0, 200){\includegraphics[angle=-90,scale=0.33]{fg08c.ps}}
%\put (190, 370) { (a)}
%\put (440, 370) { (b)}
%\put (190, 170) { (c)}
%\end{picture}
\caption{A sampling of the quality assurance plots output by {\tt excal} 
for the PT data taken on the night of MJD 53501: 
%(a) photometric zeropoints ($a$ coefficients) vs.\ time for each of the five passbands; 
(a) first-order extinctions ($k$ coefficients) vs.\ time
for each of the five passbands, and (b) magnitude residuals for the
$u'g'r'i'z'$ standard stars observed over the course of the night (the
zero lines are staggered by 0.1mag offsets in order to include the
residuals from all five filters on the same plot).  Note that, in (b),
solid symbols denote the observations that were included in the
solution, whereas open symbols denote the observations that were
removed from the solution.  
\label{excalQAa}} 
\end{figure*}

%\clearpage 
\begin{figure*}
\begin{picture}(400,600)
\put ( 50, 625){\includegraphics[angle=-90,scale=0.55]{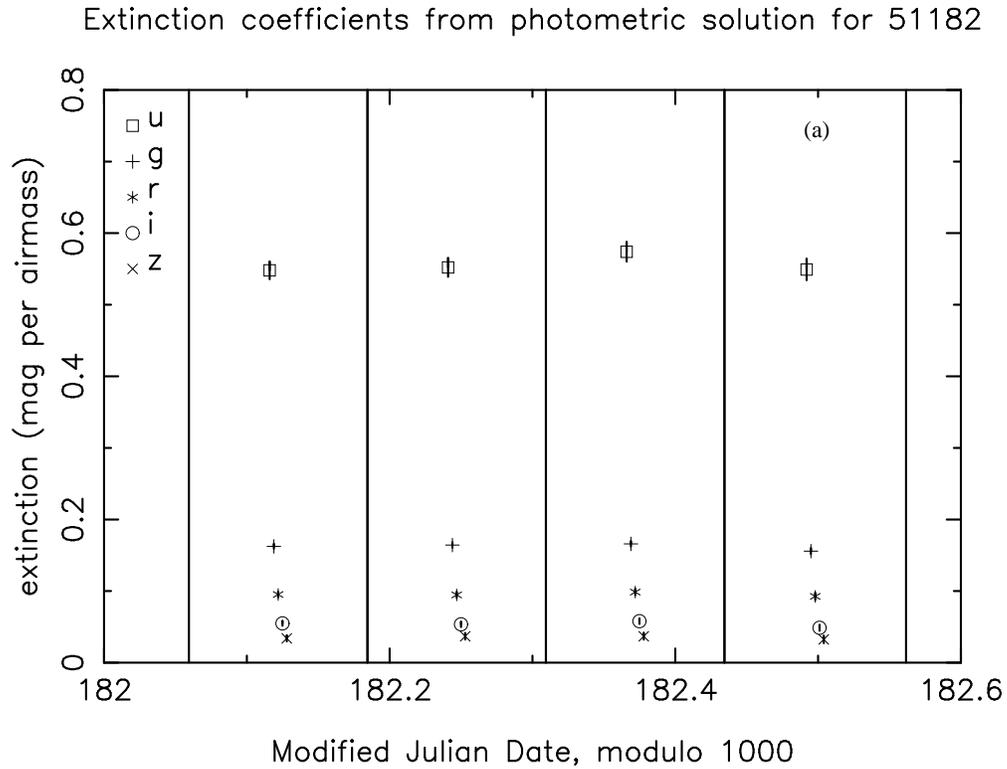}}
\put ( 50, 300){\includegraphics[angle=-90,scale=0.55]{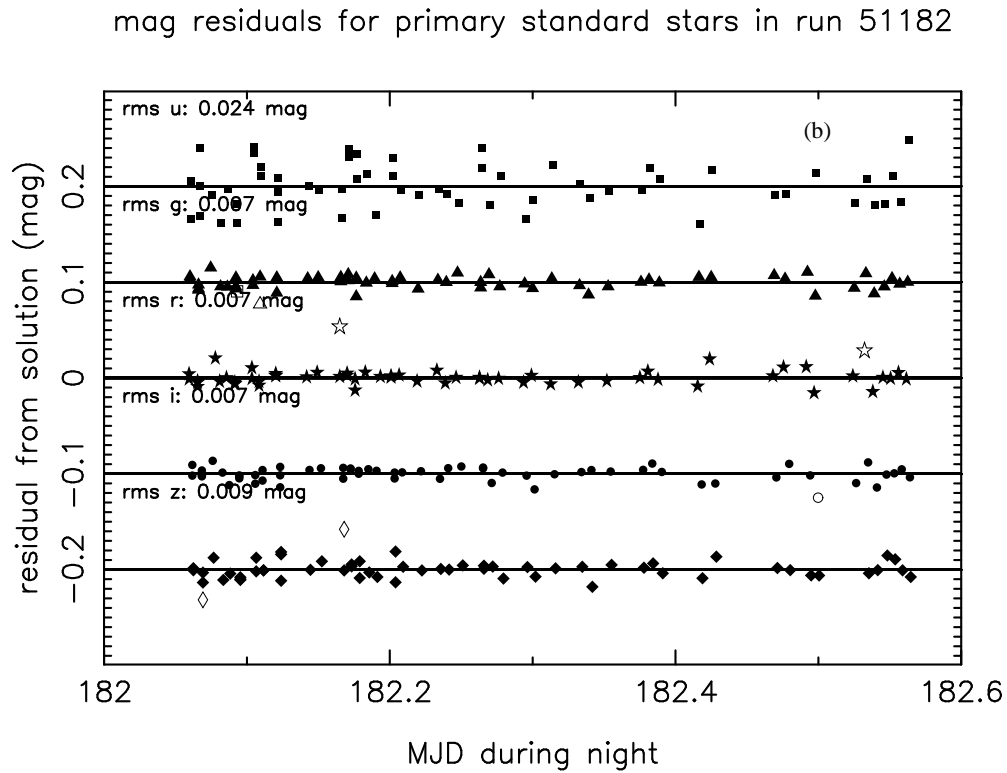}}
\put (360, 575) { (a)}
\put (360, 250) { (b)}
\end{picture}
%\begin{picture}(400,400)
%\put (  0, 400){\includegraphics[angle=-90,scale=0.33]{fg09a.ps}}
%\put (250, 400){\includegraphics[angle=-90,scale=0.33]{fg09b.ps}}
%\put (  0, 200){\includegraphics[angle=-90,scale=0.33]{fg09c.ps}}
%\put (190, 370) { (a)}
%\put (440, 370) { (b)}
%\put (190, 170) { (c)}
%\end{picture}
\caption{The same {\tt excal} quality assurance plots as in Figure~\ref{excalQAa}, 
but for the USNO-1.0m data taken on the night of MJD 51182, in which the $k$ terms
were solved for in 3-hour time blocks.
\label{excalQAb}} 
\end{figure*}

\begin{table}
\centering 
\caption{PT Color Indices ({\tt excal} \& {\tt kali})}
\label{colorIndices}
\begin{tabular}{rr}\hline
filter & color index \\
       & $(x-y)_{\rm o}$ \\
\hline
$u'$  &  $(u'-g')_{\rm o}$  \\
$g'$  &  $(g'-r')_{\rm o}$  \\
$r'$  &  $(r'-i')_{\rm o}$  \\
$i'$  &  $(r'-i')_{\rm o}$  \\
$z'$  &  $(i'-z')_{\rm o}$  \\
\hline
\end{tabular}
\end{table}

\begin{table}
\centering
\caption{PT Color Zeropoints ({\tt excal} \& {\tt kali})}
\label{colorZPs}
\begin{tabular}{rr}\hline
color zeropoint & value \\
\hline
$(u'-g')_{\rm o,zp}$ & 1.39   \\
$(g'-r')_{\rm o,zp}$ & 0.53   \\
$(r'-i')_{\rm o,zp}$ & 0.21   \\
$(i'-z')_{\rm o,zp}$ & 0.09   \\
\hline
\end{tabular}
\end{table}

\begin{table}
\centering
\caption{Instrumental Color Terms For the PT ({\tt excal})}
\label{bterms}
\begin{tabular}{rrrrrr}\hline
$<$MJD$>$ & $b(u')$ & $b(g')$ & $b(r')$ & $b(i')$ & $b(z')$ \\
\hline
\multicolumn{6}{c}{Old $u'g'r'i'z'$ filters}\\
51551      &  0.001 &  0.023 &  0.024 &  0.028 &  0.002 \\
51580      &  0.001 &  0.025 &  0.025 &  0.029 &  0.002 \\
51609      &  0.001 &  0.024 &  0.024 &  0.028 &  0.002 \\
51639      &  0.001 &  0.023 &  0.024 &  0.028 &  0.002 \\
51668      &  0.001 &  0.015 &  0.022 &  0.025 &  0.002 \\
51698      &  0.001 & -0.003 &  0.017 &  0.017 &  0.002 \\
51727      &  0.001 & -0.015 &  0.014 &  0.012 &  0.002 \\
51786      &  0.001 & -0.015 &  0.014 &  0.012 &  0.002 \\
51815      &  0.001 &  0.008 &  0.020 &  0.021 &  0.002 \\
51845      &  0.001 &  0.020 &  0.023 &  0.027 &  0.002 \\
51874      &  0.001 &  0.027 &  0.025 &  0.029 &  0.002 \\
51904      &  0.001 &  0.032 &  0.026 &  0.031 &  0.002 \\
51934      &  0.001 &  0.034 &  0.027 &  0.032 &  0.002 \\
51964      &  0.001 &  0.033 &  0.027 &  0.032 &  0.002 \\
51993      &  0.001 &  0.031 &  0.026 &  0.031 &  0.002 \\
52023      &  0.001 &  0.024 &  0.024 &  0.028 &  0.002 \\
52052      &  0.001 &  0.014 &  0.022 &  0.024 &  0.002 \\
52082      &  0.001 & -0.004 &  0.017 &  0.017 &  0.002 \\
52111      &  0.001 & -0.016 &  0.014 &  0.012 &  0.002 \\
\multicolumn{6}{c}{New $u'g'r'i'z'$ filters}\\
$\ge$52140 &  0.001 & -0.041 &  0.009 &  0.010 &  0.002 \\
\hline
\end{tabular}
\end{table}

\begin{table}
\centering
\caption{PT Standard Star Color Ranges ({\tt excal} \& {\tt kali})} 
\label{colorRanges}
\begin{tabular}{r}\hline
color \\  
\hline
   $0.70 \leq (u'-g')_{\rm o} \leq 2.70$  \\
   $0.15 \leq (g'-r')_{\rm o} \leq 1.20$  \\
   $-0.10 \leq (r'-i')_{\rm o} \leq 0.60$  \\
   $-0.20 \leq (i'-z')_{\rm o} \leq 0.40$  \\
\hline
\end{tabular}
\end{table}

The third package, {\tt excal}, takes the output of {\tt mtFrames} for
the $u'g'r'i'z'$ standard star fields, identifies the individual
standard stars within these fields, and --- using the instrumental
magnitudes for these stars as input --- invokes a least squares
routine to calculate the photometric zeropoint and the atmospheric
extinction in each filter passband.  Since the overall system response
in each filter (which is measured by the photometric zeropoints) is
unlikely to change significantly over the course of single night, a
single photometric zeropoint is calculated for a night.  On the other
hand, changes in the atmospheric transparency in each filter (which is
measured by the atmospheric extinction term), can vary significantly
over the course of a single night.  Therefore, {\tt excal} has the
option to solve for the extinction in blocks of time that cover a
night.  In processing the original $u'g'r'i'z'$ standard star network,
typically 3-hour blocks were used; in standard PT operations,
typically the block size is set to 15 hours or more, in order to solve
for only a single nightly extinction in each passband.  Instrumental
color terms and second-order (color$\times$airmass) extinctions may
also solved for, although generally multiple nights of data are needed
to determine these with any confidence, so their values are usually
set to pre-determined defaults.  Uncalibrated candidate standard stars
(i.e., stars which are believed to be non-variable but do not have
previously determined $u'g'r'i'z'$ magnitudes) can also be used in the
least squares solution as extinction standards; a useful output of the
least squares routine is an estimate of their calibrated $u'g'r'i'z'$
magnitudes.

The photometric equations solved for by {\tt excal} have, for a given
filter $x$ and color index $x-y$, the following generic form:
\begin{eqnarray}
x_{\rm inst} & = & x_{\rm o} + a_{x} + k_{x} X \nonumber \\
               &   & + b_{x} [(x-y)_{\rm o} - (x-y)_{\rm o,zp}] \nonumber \\
               &   & + c_{x} [(x-y)_{\rm o} - (x-y)_{\rm o,zp}] [X-X_{\rm zp}] ,
\end{eqnarray}
where $x_{\rm inst}$ is the measured instrumental magnitude in filter
$x$, $x_{\rm o}$ is the extra-atmospheric magnitude, $(x-y)_{\rm o}$
is the extra-atmospheric color, $a_{x}$ is the nightly zero point,
$k_{x}$ is the first order extinction coefficient, $b_{x}$ is the
system transform coefficient, $c_{x}$ is the second order (color)
extinction coefficient, and $X$ is the airmass of the observation.
$X_{\rm zp}$ and $(x-y)_{\rm o,zp}$ are zeropoint constants for the
airmass $X$ and the color index $(x-y)$, respectively.

For standard PT reductions, the filter bands are linked to the color
indices shown in Table~\ref{colorIndices}, the color zeropoint
constants are set to the values shown in Table~\ref{colorZPs}, and
$X_{\rm zp}$ is set to 1.3 (roughly the average airmass of the
standard star observations).  
Furthermore, since their values are all very small and since removing
them simplifies the fits to the data, the second-order extinction
coefficients (the $c_x$'s) are all set to zero for standard PT
reductions.

Note that the choice of which color index is linked with which filter
band is an option in {\tt excal}.  For instance, in our effort to
extend the $u'g'r'i'z'$ standard star network into the southern
hemisphere (see \S~\ref{sec:otherprojects}), we link the $i'$ band to
the $(i'-z')$ color index 
rather than to the $(r'-i')$ color index used in the standard PT
reductions, in order to maintain consistency with the photometric
equations used in establishing the original $u'g'r'i'z'$ standard star
system (Smith et al. 2002).

The photometric equations are solved iteratively, performing the
following loop:
\begin{enumerate}

\item The photometric equations are fed the current estimates for the
	extra-atmospheric magnitude, colors, and photometric coefficients
	(e.g., for the $g'$ passband, these would be $g'_{\rm o}$, $(g'-r')_{\rm o}$,
	$a_g$, $b_g$, $c_g$, and $k_g$).  For the first iteration of this
	loop, initial estimates for these parameters are fed into the
	equations. 

\item The photometric equation for each of the passbands is solved in turn.  
        This is done for all 5 passbands.  If any stars (e.g.,
        extinction standards) are having their magnitudes (e.g.,
        $g'_{\rm o}$) solved for, their colors (e.g., $(g'-r')_{\rm
        o}$) are kept fixed at the values they had at the first step
        of this loop.  (Having the colors change as their magnitudes
        in each filter are updated as one goes from one filter's
        equation to the next has been found to yield unstable
        solutions.)

\item Next, after the equations for all 5 passbands have been solved,
	if any extinction star magnitudes, are being solved for, their 
	colors must be updated and saved for the next iteration.

\item If {\tt excal} is being run interactively (a rare occurrence nowadays
	for PT processing; see \S~\ref{sec:ptfactory}), then the
	residuals from the photometric solution are displayed to the
	screen, in each passband, and the user is allowed to discard
	stars from the solution (or add back in stars previously
	discarded).

	If the user {\em does} add/remove any star to/from the
	solution, {\tt excal} returns immediately back to the top of
	the loop and starts again; otherwise, {\tt excal} exits
	out of the loop.

\item If {\tt excal} is being run non-interactively, it 
	sigma-clips the outliers in the residuals and returns to the
	top of the loop.  In this case, the number of iterations of
	this loop is controlled by an adjustable parameter.  For
	normal PT operations, the clipping is done at the 2.5$\sigma$
	level, and the loop performs 3 iterations.

\end{enumerate}

One can also choose to fix any of the photometric coefficients to a
preset value and {\em not} solve for it.  For instance, for the PT, we
set the instrumental color terms to the preset values shown in
Table~\ref{bterms} (linearly interpolating between MJDs as
needed). Note that, up through July 2001, the PT had a set of
$u'g'r'i'z$ filters which showed a fairly strong seasonal variations;
the current set of PT $u'g'r'i'z'$ filters, installed in August 2001,
have much more stable transmissions.

For PT reductions, where the prime focus is providing secondary patch
stars calibrated on the SDSS 2.5m $ugriz$ system, the set of standard
stars used in the fits to the photometric equations 
are constrained to a set of relatively narrow color ranges (shown in
Table~\ref{colorRanges}).  Only standard stars that fall within all
four of these color ranges are used, for, within these narrow ranges,
the transformation from the USNO $u'g'r'i'z'$ system to the SDSS
$ugriz$ system is best determined and known to be very linear.

Finally, {\tt excal}, like {\tt mtFrames}, also outputs quality
assurance plots for the night processed.  Figure~\ref{excalQAa} shows
quality plots for a night of PT data (MJD 53501) in which the
first-order extinctions ($k$-terms) were solved for in one long
(15-hour) time block that covers the whole night.
Figure~\ref{excalQAb} shows quality plots for a night of USNO-1.0m
data (MJD 51182) in which the $k$ first-order extinctions where solved
for in shorter (3-hour) time blocks.

Figures~\ref{excalQAa}a and \ref{excalQAb}a show the first-order
extinctions for the two nights in question; 
Figures~\ref{excalQAa}b and \ref{excalQAb}b, the residuals all of the
standard stars observed over the course of each of these two nights.
For PT reductions, we consider a night photometric if the rms
residuals are $\sigma_{u'} \leq 0.040$~mag, $\sigma_{g'} \leq
0.025$~mag, $\sigma_{r'} \leq 0.025$~mag, $\sigma_{i'} \leq
0.025$~mag, {\em and} $\sigma_{z'} \leq 0.040$~mag.

\subsection{\tt kali}

%\clearpage 
\begin{figure*}
\includegraphics[angle=-90,scale=0.7]{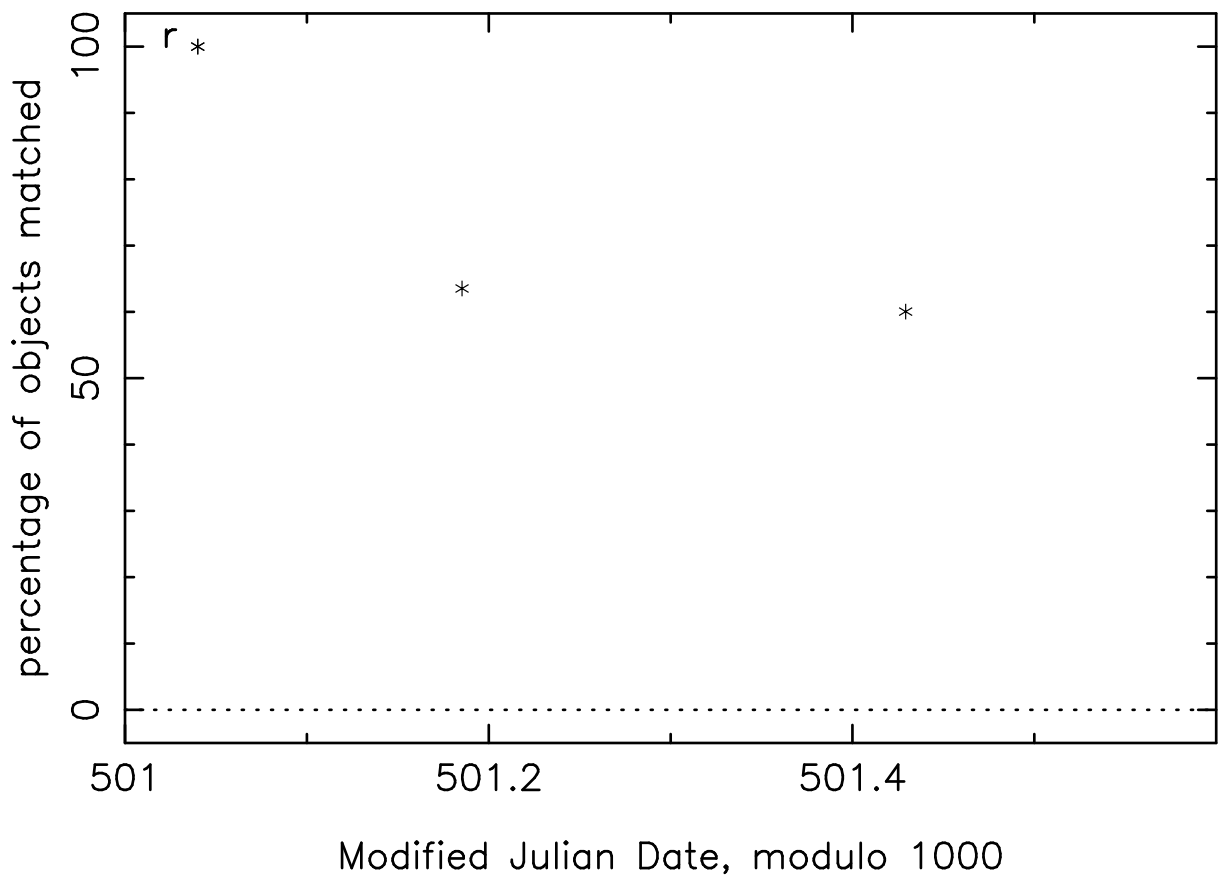}
%\begin{picture}(400,400)
%\put (  0,400){\includegraphics[angle=-90,scale=0.33]{fg10a.ps}}
%\put (250,400){\includegraphics[angle=-90,scale=0.33]{fg10b.ps}}
%\put (  0,200){\includegraphics[angle=-90,scale=0.33]{fg10c.ps}}
%\put (250,200){\includegraphics[angle=-90,scale=0.33]{fg10d.ps}}
%\put (200, 360) { (a)}
%\put (450, 360) { (b)}
%\put (200, 160) { (c)}
%\put (450, 160) { (d)}
%\end{picture}
%\caption{A sampling of the quality assurance plots output by {\tt kali} 
%for the PT data taken on the night of MJD 53501: (a) the percentage of
%stars detected in the $r'$-band matched to stars in the finding chart
%file for each of the secondary patches and manual target fields
%observed during the night (note that the finding chart file for each
%secondary patch and manual target field is extracted from the the
%Guide Star Catalog-I; Lasker et al.\ 1990); (b) the combined $g-r$ vs.\
%$u-g$ color-color diagram, (c) the combined $r-i$ vs. $g-r$
%color-color diagram, and (c) the combined $i-z$ vs. $r-i$ color-color
%diagram for the stars in all the secondary patch and manual fields
%from the night.  The lines in (b) and (c) show the $u'g'r'i'z'$
%stellar loci (based on the stellar loci of Lenz et al.\ 1998).
\caption{One of the quality assurance plots output by {\tt kali} 
for the PT data taken on the night of MJD 53501:  the percentage of
stars detected in the $r'$-band matched to stars in the finding chart
file for each of the secondary patches and manual target fields
observed during the night (note that the finding chart file for each
secondary patch and manual target field is extracted from the the
Guide Star Catalog-I; Lasker et al.\ 1990).
\label{kaliQA_1}} 
\end{figure*}

%\clearpage 
\begin{figure*}
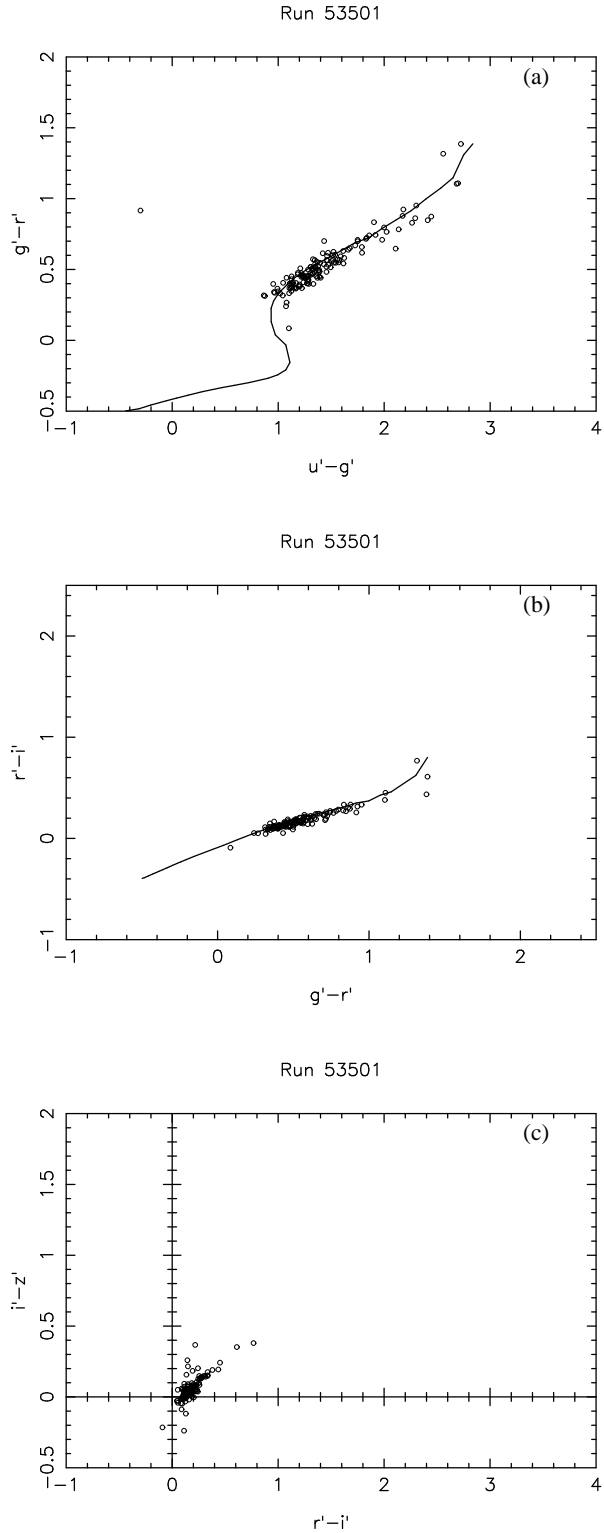

\begin{picture}(400,600)
\put (100,600){\includegraphics[angle=-90,scale=0.34]{fg11a.ps}}
\put (100,400){\includegraphics[angle=-90,scale=0.34]{fg11b.ps}}
\put (100,200){\includegraphics[angle=-90,scale=0.34]{fg11c.ps}}
\put (300, 570) { (a)}
\put (300, 370) { (b)}
\put (300, 170) { (c)}
\end{picture}
\caption{Color-color quality assurance plots output by {\tt kali} 
for the PT data taken on the night of MJD 53501: (a) the combined $g-r$ vs.\
$u-g$ color-color diagram, (b) the combined $r-i$ vs. $g-r$
color-color diagram, and (c) the combined $i-z$ vs. $r-i$ color-color
diagram for the stars in all the secondary patch and manual fields
from the night.  The lines in (a) and (b) show the $u'g'r'i'z'$
stellar loci (based on the stellar loci of Lenz et al.\ 1998).
\label{kaliQA_2}} 
\end{figure*}

The last package used in normal operations, {\tt kali}, performs the
astrometric calibration of the secondary and applies the
photometric solutions produced by {\tt excal} to the secondary patch
instrumental magnitudes.  As noted above, these secondary patch
fields, which are observed by the PT, are later scanned over by the
imaging camera on the SDSS 2.5m telescope.  Thus, they act as
calibration transfer fields to set the photometric zeropoints for the
imaging camera data.  The {\tt kali} package can also be applied to
calibrate manual, or target-of-opportunity, fields observed during a
night.

Astrometry is performed using a triangle matching technique based upon
the one used by the Faint-Object Classfication and Analysis System
(FOCAS; Valdes et al.\ 1995).\footnote{A standalone ANSI~C version of
this code can be obtained from {\tt http://spiff.rit.edu/match/}.} For
standard PT processing, finding charts extracted from the Guide Star
Catalog 1.1 (Lasker et al.\ 1990) are used for the matching. For telescopes
with smaller fields-of-view (e.g., the USNO-1.0m and the CTIO-0.9m;
see \S~\ref{sec:otherprojects}), however, the deeper (and hence more
densely populated) Guide Star Catlogue 2.2\footnote{{\tt
http://www-gsss.stsci.edu/gsc/gsc2/GSC2home.htm}} is sometimes
necessary to obtain a good astrometric solution for a given field.

{\tt kali} outputs the photometry for the secondary patches both in
the USNO $u'g'r'i'z'$ system and in the SDSS $ugriz$ magnitudes.  {\tt
kali} first applies the $u'g'r'i'z'$ photometric solutions from {\tt
excal} to the instrumental magnitudes of the secondary patches.  This
step is performed iteratively, until the calibrated $u'g'r'i'z'$
magnitudes have converged. To convert from the USNO $u'g'r'i'z$ system
to the SDSS 2.5m telescope's $ugriz$ system, {\tt kali} applies the
following transformation equations:
\begin{eqnarray}
  u_{\rm o} & = & u'_{\rm o} \\
  g_{\rm o} & = & g'_{\rm o} + 0.060 \times [ (g'-r')_{\rm o} - 0.53 ]  \\
  r_{\rm o} & = & r'_{\rm o} + 0.035 \times [ (r'-i')_{\rm o} - 0.21 ]  \\
  i_{\rm o} & = & i'_{\rm o} + 0.041 \times [ (r'-i')_{\rm o} - 0.21 ]  \\
  z_{\rm o} & = & z'_{\rm o} - 0.030 \times [ (i'-z')_{\rm o} - 0.09 ]  . 
\end{eqnarray}

Strictly speaking, the SDSS $ugriz$ magnitudes output by {\tt kali}
only apply within the narrow color ranges described by
Table~\ref{colorRanges}, although they are generally accurate for a
much broader range, at least for normal stars (e.g., stars showing no
strong emission features in the their spectra) blueward of the M0
spectral type (see, for example, Fig.~7 and 8 of Rider et al.\ 2004).

Further, like {\tt mtFrames} and {\tt excal}, {\tt kali} outputs some
quality assurance plots for the night processed, a sampling of which
is shown in Figures~\ref{kaliQA_1} and \ref{kaliQA_2}.
Figure~\ref{kaliQA_1} shows how well the PT patches were matched to
the the stars in the finding charts, and Figures \ref{kaliQA_2}a, b,
and c show the $ugr$, $gri$, and $irz$ color-color diagrams,
respectively, for the stars in the secondary patches observed on this
night (the $ugr$ and $gri$ diagrams also show a stellar locus (Lenz et
al.\ 1998) for comparison).

Finally, {\tt kali} automatically tags the quality of the photometry
within individual secondary and manual patches by monitoring the
position of certain features in each patch's color-color diagrams.
For determining the quality of the photometry in the $griz$ bands,
{\tt kali} uses the location in $g'-r'$, $r'-i'$, $i'-z'$ space where
the red and blue branches of the stellar locus cross --- i.e., roughly
where stars of spectral type M0 sit, which is at $g'-r' \approx 1.35$,
$r'-i' \approx 0.50$, and $i'-z' \approx 0.25$; this is the so-called
Zhed point test, a more refined version of which is described in
detail by Ivezi\'{c} et al.\ (2004).  The quality of the $u$ band photometry is
automatically tagged based upon the estimated $u'-g'$ color of stars
in the blue branch at $g'-r'=0.50$, which is $u'-g'\approx 1.45$.

\subsection{Auxiliary Packages}

In addition to the above four packages, which are used for normal
night-to-night survey operations, there are two other packages ---
{\tt solve\_network} and {\tt superExcal} --- which were developed and
used for setting up the $u'g'r'i'z'$ primary standard star network of 
Smith et al.\ (2002).  These two packages are similar to {\tt excal}, in
that they invoke least squares routines to solve for the photometric
parameters of a set of data and for the best-fit magnitudes of the
primary standards.  They differ from {\tt excal} in that, whereas the
main purpose of {\tt excal} is the standard, single-night solution of
the photometric equations during routine survey operations, the main
purpose of {\tt solve\_network} and {\tt superExcal} was to calibrate
the SDSS standard star network.  Both {\tt solve\_network} and {\tt
superExcal} use a single star to set the zeropoint for the photometric
solution.  For setting up the SDSS standard star network, this star
was the F subdwarf BD+17$^{\circ}$4708, whose magnitudes were {\em
defined} to be $u'=10.56$, $g'=9.64$, $r'=9.35$. $i'=9.25$, $z'=9.23$,
based upon the spectrophotometric calculations of Fukugita et al.\ (1996).
Thus, BD+17$^{\circ}$4708 sets the zeropoint for the entire standard
star network.

Although they share the same goal and the same basic methodology, {\tt
solve\_network} and {\tt superExcal} do differ in some important
respects, the most important being that the code for {\tt
solve\_network} was constructed independently of that for {\tt excal},
whereas {\tt superExcal} was basically an outgrowth of {\tt excal}.
As a result, the separate results from these two different packages
provide useful cross-checks on the final calibrations of the standard
star system.  Results and output of {\tt solve\_network} and {\tt
superExcal} can be found in Smith et al.\ (2002).

Finally, although not strictly an auxiliary package but more of an
auxiliary mode of operations is the {\tt MTPIPE} follow mode, which
permits the near-realtime running of {\tt mtFrames} and {\tt excal}
during data acquisition at the PT.  {\tt MTPIPE} in follow mode was
used by the SDSS observers during the 2000--2001 time frame in
conjunction with the {\tt hoggpt} photometricity monitor
(Hogg et al.\ 2001) in order to track sky conditions over the course of a
night.  This functionality is nowadays completely handled by the
dedicated {\tt hoggpt} photometricity monitor.

\section{Tests of {\tt MTPIPE}}
\label{sec:tests}

%\clearpage 
\begin{figure*}
\includegraphics[angle=0,scale=0.75]{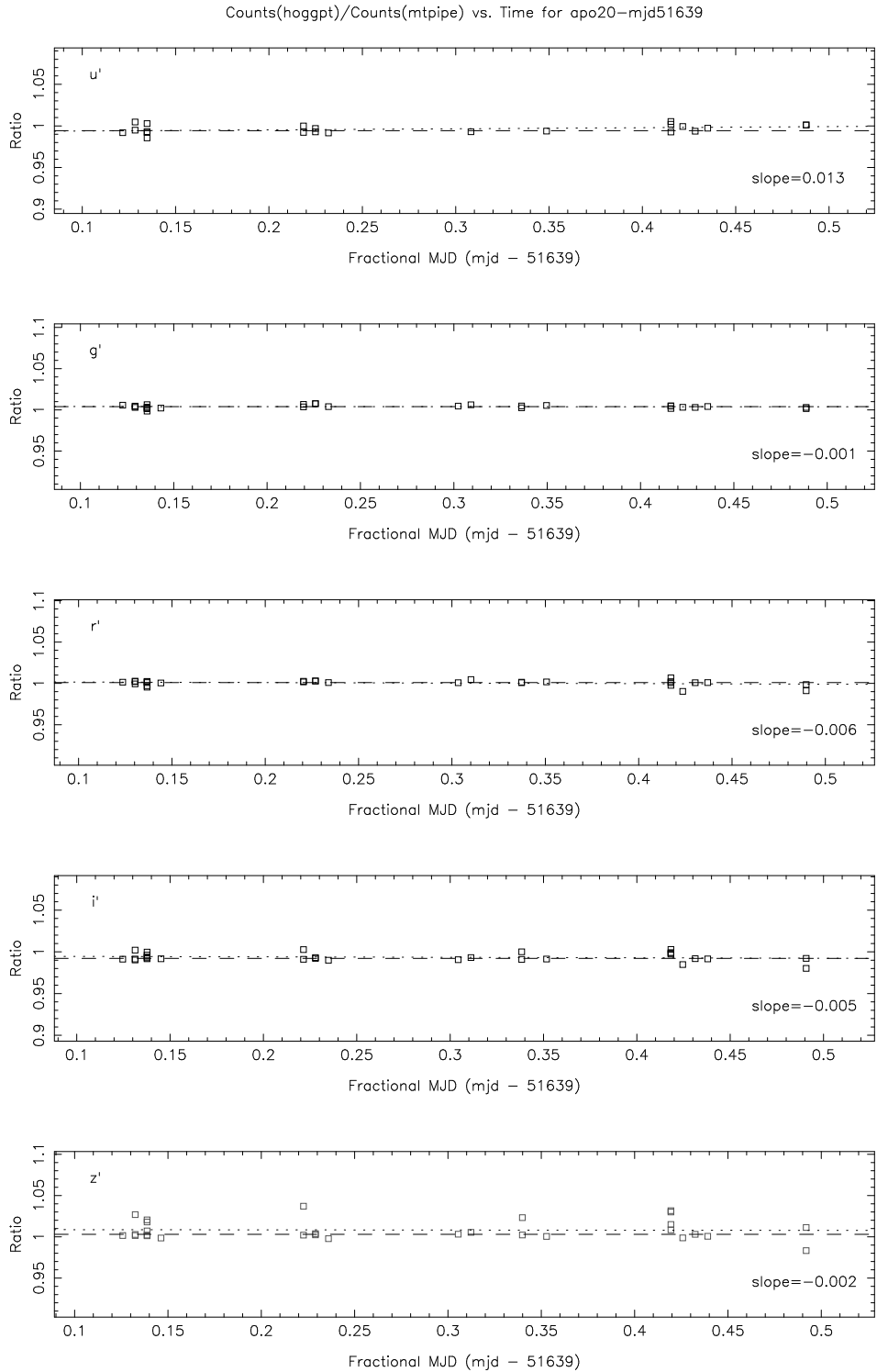} 
\caption{Comparison of aperture counts for standard stars from {\tt hoggpt} 
and from {\tt mtFrames} over the course of the night for the PT data
from the night of MJD 51639.  The dashed line in each panel is the
median of the ratio of Counts({\tt hoggpt})/Counts({\tt MTPIPE}) for
data; its slope is flat.  The dotted line in each panel is the best
fit line to the data; its slope is noted in the panel's legend.
\label{compareHoggptWithMtpipeInstrMags1}} 
\end{figure*}

%\clearpage 
\begin{figure*}
\includegraphics[angle=0,scale=0.75]{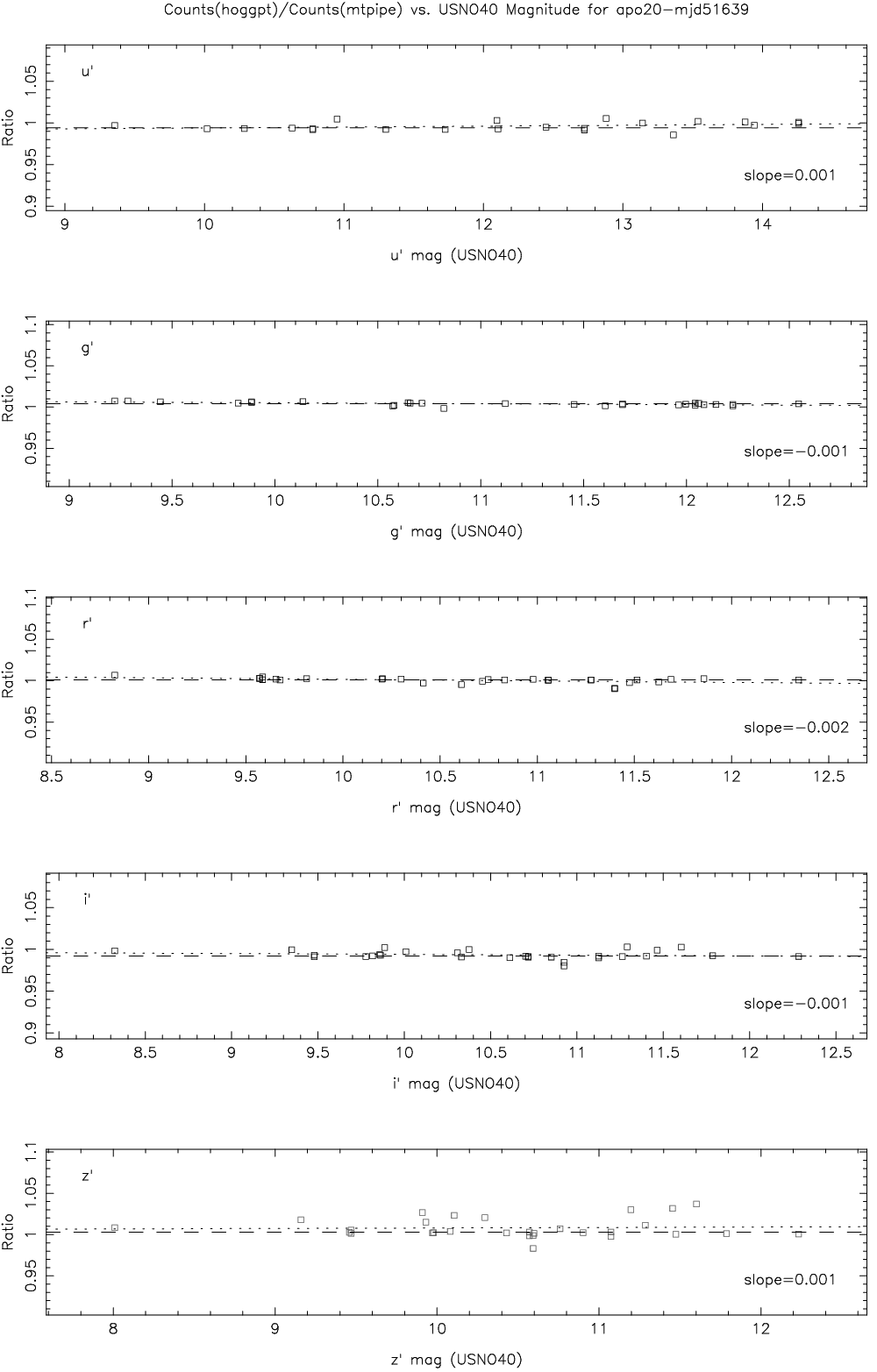} 
\caption{Comparison of aperture counts for the Smith et al.\ (2002) standard stars from {\tt hoggpt} 
and from {\tt mtFrames} vs.\ the USNO standard magnitude for the PT
data from the night of MJD 51639.  Dashed and dotted lines are as in
Figure~\ref{compareHoggptWithMtpipeInstrMags1}.
\label{compareHoggptWithMtpipeInstrMags2}} 
\end{figure*}

%\clearpage 
\begin{figure*}
\includegraphics[angle=0,scale=0.75]{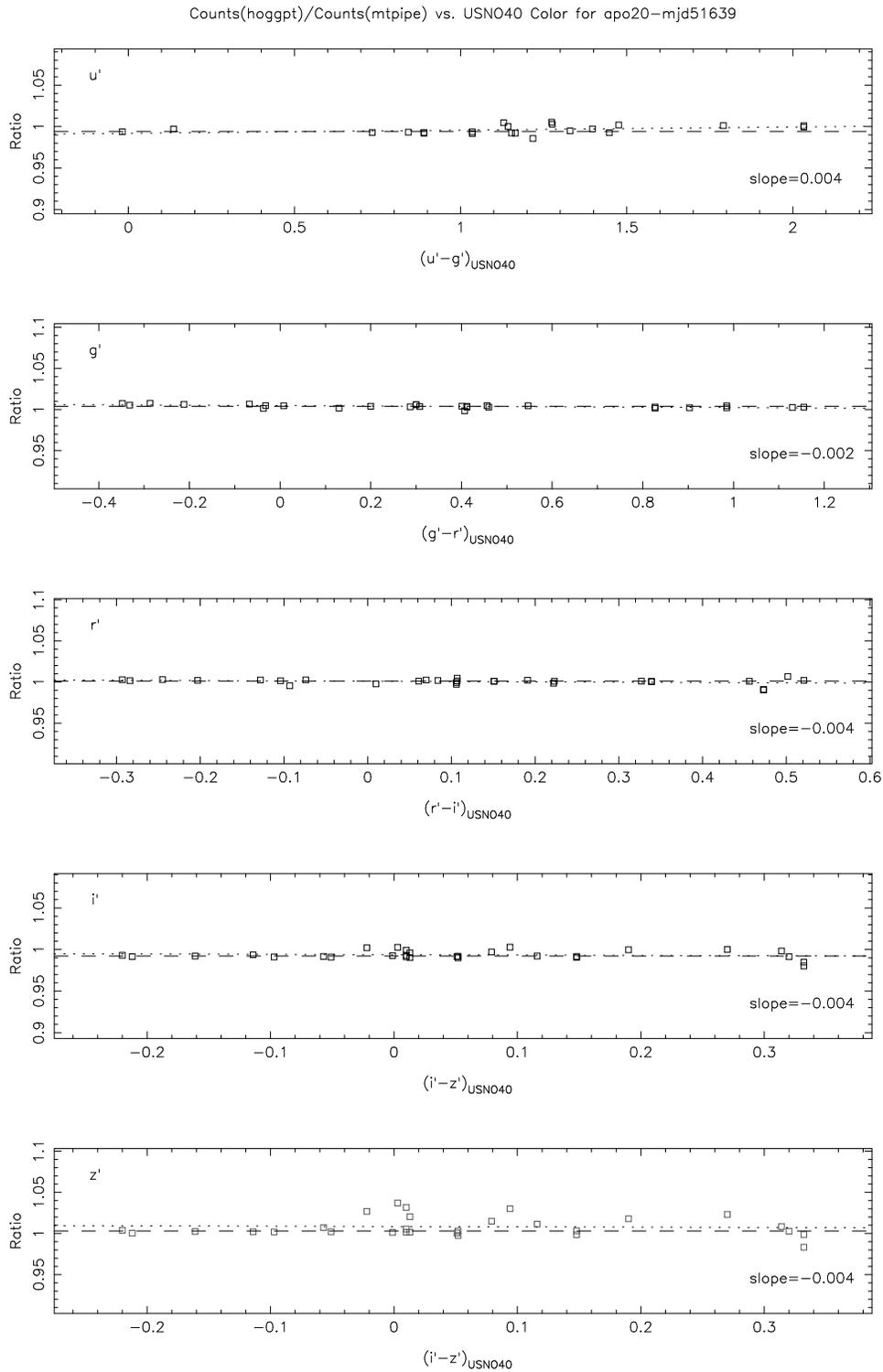} 
\caption{Comparison of aperture counts for the Smith et al.\ (2002) standard stars from {\tt hoggpt} 
and from {\tt mtFrames} vs.\ the USNO standard color for the PT
data from the night of MJD 51639.  Dashed and dotted lines are as in
Figure~\ref{compareHoggptWithMtpipeInstrMags1}.
\label{compareHoggptWithMtpipeInstrMags4}} 
\end{figure*}

%\clearpage 
\begin{figure*}
\includegraphics[angle=0,scale=0.75]{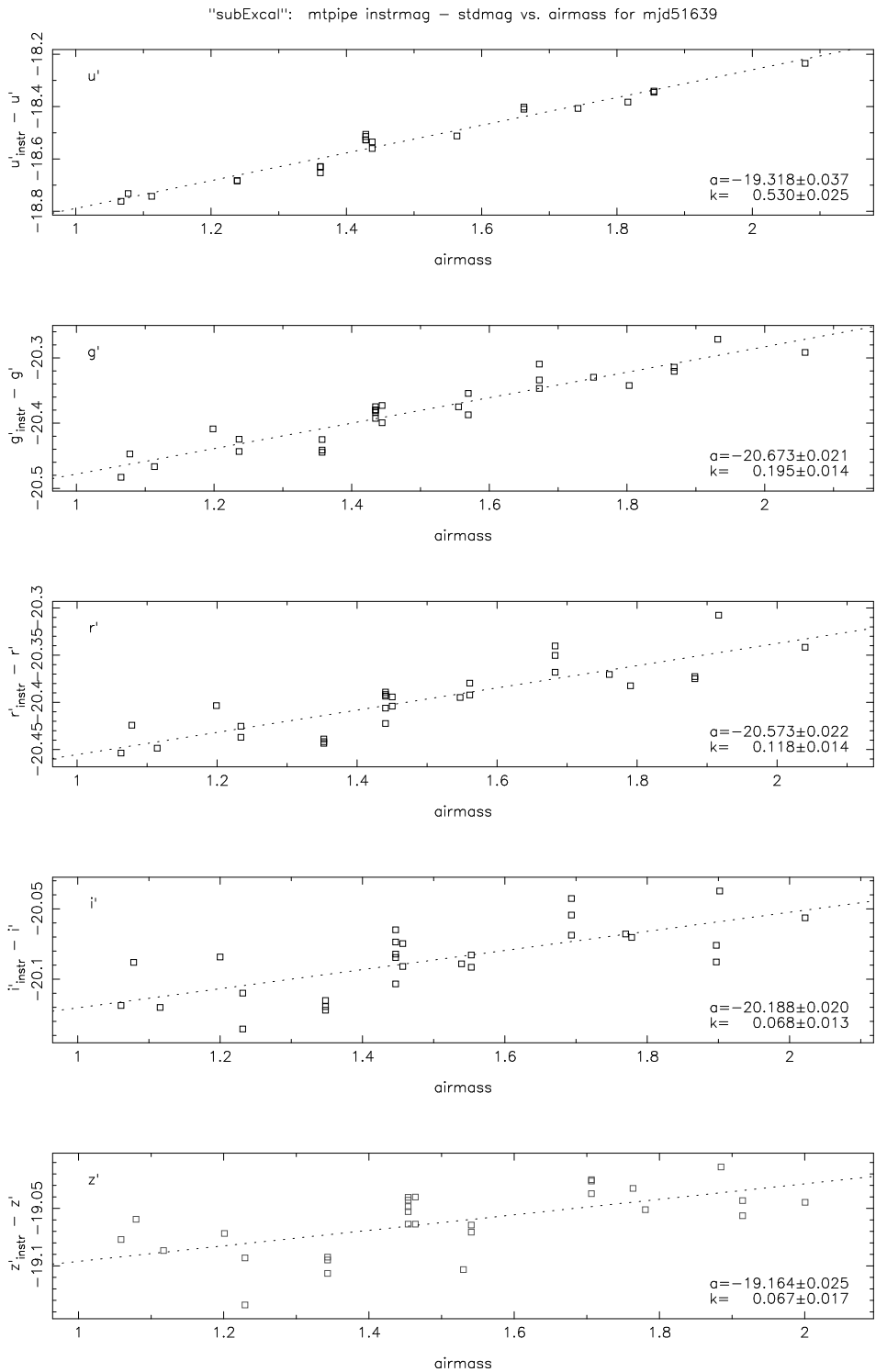}
\caption{``{\tt subExcal}'' results for photometric zeropoints ($a$ terms) and
first-order extinctions ($k$ terms) for the PT data from the night of MJD 51639.
The dotted line in each panel shows the best fit line to the data.
The values for the slope ($k$) and zeropoint ($a$) of the best fit line
is shown in each panel's legend. 
\label{subExcalak}} 
\end{figure*}

%\clearpage 
\begin{figure*}
\includegraphics[angle=0,scale=0.75]{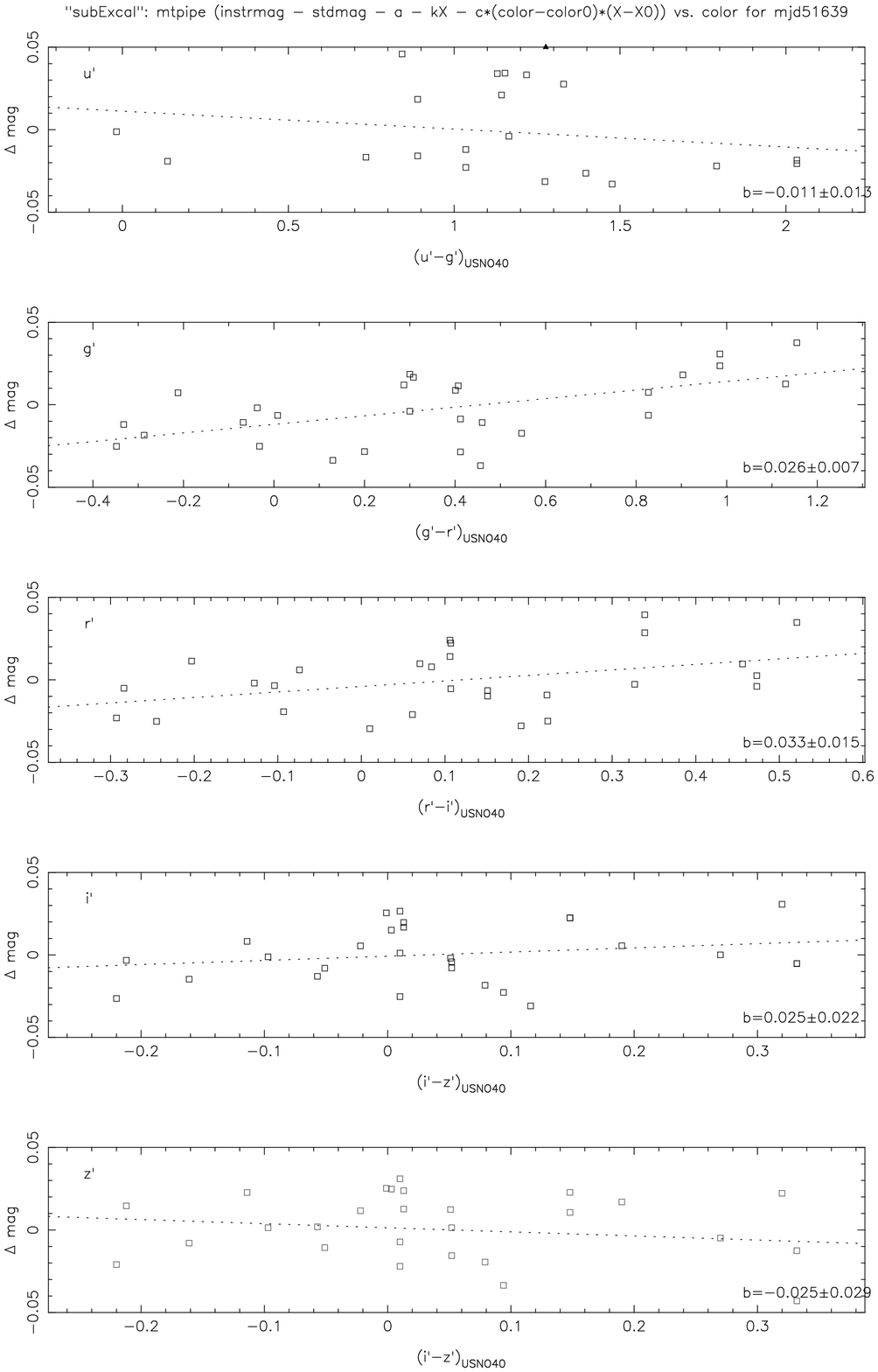}
\caption{``{\tt subExcal}'' results for instrumental color ($b$) term coefficients 
for the PT data from the night of MJD 51639.  The dotted line in each
panel shows the best fit line to the data.  The value for the
slope ($b$) of the best fit line is shown in each panel's legend.
\label{subExcalb}} 
\end{figure*}

As a critical component in the photometric calibration of the SDSS,
{\tt MTPIPE} has undergone a variety of tests to ensure the validity
of its outputs.  Most of these tests have been aimed at the
potentially trickier parts of {\tt MTPIPE} --- the image processing
and aperture photometry in {\tt mtFrames} and the photometric
equations solver in {\tt excal}.

A major test of {\tt mtFrames} was accomplished by comparing its
outputs with those of the {\tt hoggpt} APO photometricity monitor
(Hogg et al. 2001) for various nights of PT data.  The {\tt hoggpt}
software was written completely independently of {\tt mtpipe} in the
data analysis language IDL.\footnote{{\tt http://www.rsinc.com/}} It
thus could act as an independent check of the data processing methods
utilized in {\tt mtFrames}.  Overall, it was discovered that these
pipelines created virtually identical master bias frames, master flat
fields, and master fringes.  Furthermore, for a given aperture size,
the two pipelines yielded essentially indistinguishable aperture
photometry.  Plots comparing the aperture photometry from the two
pipelines are shown in
Figures~\ref{compareHoggptWithMtpipeInstrMags1},
\ref{compareHoggptWithMtpipeInstrMags2}, and
\ref{compareHoggptWithMtpipeInstrMags4}.  In these plots, an aperture
size of 7.43\arcsec radius is used on PT data from MJD 51639.

The basis of the {\tt excal} photometric equations solver is an
iterative non-linear matrix inversion routine which simultaneously
solves for all parameters in each filter's photometric equation.  To
test this, a very stripped-down version of {\tt excal} (``{\tt
subExcal}'') was written which utilized a vanilla linear least squares
line-fitting routine based upon the {\tt LINFIT} routine of
Bevington (1969) converted into the SDSS Tcl/{\tt ASTROTOOLS}
environment. {\tt subExcal} does not solve for all parameters
simultaneously, but first solves for the photometric zeropoint and
the first order extinction (the $a$ and $k$ coefficients) by fitting a
straight line to an equation of the form (taking the $g'$ filter as
our example):
\begin{equation}
g'_{\rm inst} - g'_{\rm o} = a_g + k_gX , 
\end{equation}
(see Fig.~\ref{subExcalak}), and then uses the values of the $a$ and $k$
terms thus obtained to solve for the $b$ coefficient by
fitting another straight line to an equation of the form:
\begin{equation}
g'_{\rm inst} - g'_{\rm o} - a_g - k_gX = b_g(g'-r')
\end{equation}
(see Fig.~\ref{subExcalb}).  In Table~\ref{excalVsSubExcal}, we
compare the output of {\tt excal} against the output of {\tt subExcal}
for PT data from the night of MJD 51639.  Note that in many cases, the
results are indistinguishable, and in all cases the values differ by
no more than 1$\sigma$.

The {\tt kali} package of {\tt MTPIPE} is more or less
straightforward.  It performs astrometry on secondary patches, and the
astrometry needs to be accurate enough for PT secondary patch stars to
be cross-matched with stars in the SDSS 2.5m imaging camera data.
Residuals in the astrometric solutions (based against the Guide Star
Catalog-I; Lasker et al.\ 1990) are typically less than $1\arcsec$~RMS in both
RA$\times$cos(DEC) and DEC, which is sufficient for the PT-2.5m
cross-matches.  Further, {\tt kali} applies the photometric solutions
from {\tt excal} to the instrumental magnitudes of the secondary
patches.  This has been verified by manually applying the {\tt excal}
solutions to the instrumental magnitudes of the PT secondary patch
data and comparing the results; the resulting differences are
typically of the order of the roundoff error of the
calculator/computer used for the manual computation. 

Of course, in the final analysis, the true-end-to-end test of {\tt
MTPIPE} is the resulting quality of the SDSS photometric calibrations.
As stated in the introduction of this paper, the rms errors measured
over the several thousand square degrees that currently make up the
SDSS imaging data set are $\approx$0.02~mag (2\%) rms over several
thousand square degrees of sky in $g$, $r$, and $i$, and
$\approx$0.03~mag (3\%) in $u$ and $z$ (Ivezi\'{c} et al.\ 2004; Adelman-McCarthy et al.\ 2006).

In the future (\S~\ref{sec:otherprojects}), {\tt MTPIPE} will likely need
to calibrate secondary patches at low Galactic latitudes for SDSS-II.
How close to the Galactic Plane (i.e, at what stellar densities) can
{\tt MTPIPE} --- which does aperture, not PSF, photometry ---
accurately calibrate these patches?  One of us (CTR) is currently
testing {\tt MTPIPE} against {\tt DAOPHOT} (Stetson 1987)  for
some low-Galactic-latitude PT fields.

%Info from /data/dp40.c/data/mt/mttest/dtucker/mtpipevshoggpt/mjd51639
% exPhoto-51639.b1.fit and photomdata-51639.b1.fit.
\begin{table}
\centering
\caption{Comparison of {\tt excal} and ``{\tt subExcal}'' for MJD 51639}
\label{excalVsSubExcal}
\tiny
\begin{tabular}{rrrrr}\hline
Filter & Method & $a_{\rm filter}$ & $b_{\rm filter}$ & $k_{\rm filter}$ \\
\hline
     &                &                     &                    &                  \\
$u'$ & {\tt excal}    &  $-19.286\pm0.042$  &  $-0.011\pm0.014$  &  $0.516\pm0.025$ \\
     & {\tt subExcal} &  $-19.318\pm0.037$  &  $-0.011\pm0.013$  &  $0.530\pm0.025$ \\
     &                &                     &                    &                  \\
$g'$ & {\tt excal}    &  $-20.668\pm0.019$  &  $+0.026\pm0.007$  &  $0.184\pm0.012$ \\
     & {\tt subExcal} &  $-20.673\pm0.021$  &  $+0.026\pm0.007$  &  $0.195\pm0.014$ \\
     &                &                     &                    &                  \\
$r'$ & {\tt excal}    &  $-20.578\pm0.021$  &  $+0.033\pm0.015$  &  $0.119\pm0.013$ \\
     & {\tt subExcal} &  $-20.573\pm0.022$  &  $+0.033\pm0.015$  &  $0.118\pm0.014$ \\
     &                &                     &                    &                  \\
$i'$ & {\tt excal}    &  $-20.194\pm0.020$  &  $+0.026\pm0.023$  &  $0.071\pm0.013$ \\
     & {\tt subExcal} &  $-20.188\pm0.020$  &  $+0.025\pm0.022$  &  $0.068\pm0.013$ \\
     &                &                     &                    &                  \\
$z'$ & {\tt excal}    &  $-19.162\pm0.026$  &  $-0.025\pm0.029$  &  $0.067\pm0.017$ \\
     & {\tt subExcal} &  $-19.164\pm0.025$  &  $-0.025\pm0.029$  &  $0.067\pm0.017$ \\
     &                &                     &                    &                  \\
\hline
\end{tabular}
\end{table}

\section{The PT Factory}
\label{sec:ptfactory}

For the photometric calibration of the SDSS 2.5m imaging camera data,
a method is needed for the timely, end-to-end processing of the PT
data that is fully integrated and maintained within the official SDSS
data processing environment at Fermilab.  This need is fulfilled by
the PT Factory, a term which applies to a suite of scripts that
oversees the day-to-day processing and quality assurance of PT data at
Fermilab.  These scripts reside in the {\tt dp} software data
processing product, which is a collection of UNIX shell scripts and
Tcl/{\tt ASTROTOOLS} code that oversees the overall processing and
quality assurance for 2.5m imaging, 2.5m spectroscopy, and PT data
(Stoughton et al.\ 2002b).

Each morning, the PT Factory performs an {\tt rsync}\footnote{{\tt
http://rsync.samba.org/}} copy of the previous night's PT data from
APO to a local disk at Fermilab (typically $\sim$2~GB for a
photometric night).  The morning {\tt rsync} also copies the night's
images and log from the APO 10 micron All-Sky Camera\footnote{{\tt
http://hoggpt.apo.nmsu.edu/irsc/tonight/}} (Hogg et al. 2001); the log,
which is a record of the sky rms for each all-sky camera image over
the course of a night, is used by the PT Factory to flag and exclude
from {\tt mtpipe} processing any PT target images afflicted by clouds.

Once the PT Factory determines that the night's PT data have been
completely transferred (typically by the early- to mid-afternoon of
the same day), {\tt MTPIPE} is automatically run on the data.  Once
{\tt MTPIPE} has run to completion, there is a manual inspection step,
in which a human checks over the quality assurance plots generated by
{\tt mtFrames}, {\tt excal}, and {\tt kali}.  This manual inspection
step exists to ensure that nothing unexpected occurred with either the
raw data or with the data processing that was not already caught by
{\tt MTPIPE}'s automated quality assurance tests.  If, as is generally
the case, the night's processing passes the manual inspection step,
the photometric solutions and survey-quality secondary patches are
copied to a location on disk where {\tt NFCALIB} expects to find them,
and the contents of the night's reductions are loaded into a database
and saved to a tape robot.

An additional input into the PT factory is the ``ptlog'' observer's
report, which is e-mailed at the end of each night to a ptlog mailing
list.  This report contains summary comments on the night, including
the sky conditions, which (if any) manual or special targets were
observed, any issues with the PT hardware or observing software, and
any other comments that might be relevant to either the PT data
processing or to the PT operations.
The ptlog is read each morning, and its contents are used in deciding
whether any manual intervention of the PT Factory is needed for that
night's data.  As such, it provides an essential link between the
observers at APO and the data processors at FNAL.

A final step of PT processing, not strictly part of the PT Factory, is
updating the secondary patch database maintained at APO.  This
database keeps track of which secondary patches have been observed,
which have not been observed, and which of the observed patches have
been determined by the PT Factory to be of survey quality.  It also
keeps track of the observing priority of the secondary patches.  This
database therefore serves as essential feedback to the observers
regarding the secondary patch observing priorities for a given dark
run or even a given night.

\section{Other SDSS-related Projects using {\tt MTPIPE}}
\label{sec:otherprojects}

Since the inception of the SDSS observing effort on the PT, we were
aware that {\tt MTPIPE} would enable additional projects to be
undertaken.  Initially, these were centered on use of the PT during
bright time, but later it became apparent that other telescopes could
be used with modification of the {\tt preMtFrames} code and creation
of additional input parameter files.

Early in the SDSS we recognized the usefulness of open star clusters
as potential calibration fields for the main survey, as well as being
good science targets in their own right.  This led to the development
of a star cluster project in the $u'g'r'i'z'$ filter system.  Also,
shortly after the establishment of the original $u'g'r'i'z'$ standard
star network (Smith et al.\ 2002), we began to receive requests for
southern hemisphere standard stars.  These requests led us to develop
a southern $u'g'r'i'z'$ standard stars project which was undertaken as
an NOAO Survey Program.  Further, as we reached the point where we
could see the end of the SDSS proper, development of survey extension
strategies were pursued by the collaboration.  The PT and the {\tt
MTPIPE} software are integral to the support of these extension
efforts.

In the rest of this section, we will briefly discuss each of the
project areas that are enabled by this unique calibration software,
and point to references for additional information.

\subsection{Southern $u'g'r'i'z'$ Standards Project}

Due to the nature and location of the SDSS, the bulk of the initial set
of standard stars to calibrate the survey were placed in the northern 
hemisphere and on the equator (Fig.~\ref{allprifieldlocationsRaDec}).
As part of the NOAO Surveys Program,\footnote{{\tt
http://www.noao.edu/gateway/surveys/programs.html}} a group of us was
granted time on the CTIO-0.9m over the 2000-2004 time period in order
to establish a network of $u'g'r'i'z'$ standard stars in the southern
hemisphere (see Fig.~\ref{allprifieldlocationsRaDec}).  Processing and
analysis of these data are nearly complete, and we are targetting the
second-half of 2006 for release of the bulk of the data.  (An initial
data release for standard stars in the Chandra Deep Field South
appeared in Smith et al.\ 2003.)  As the processing and analysis is completed,
results will be made available on our publicly accessible
website.\footnote{{\tt http://www-star.fnal.gov/}}

\subsection{$u'g'r'i'z'$ Open Clusters Project}

One of the most powerful tools available in astrophysics for exploring
and   testing theories   of   star formation,   stellar  and  galactic
evolution, and the chemical enrichment  history of  the Galaxy is  the
study of open star clusters.  
Therefore, a group of us embarked on a survey of (mostly) southern
hemisphere open star clusters using the $u'g'r'i'z'$ filter system on
small telescopes.  This project had as its original goal to
verify the SDSS calibration scheme.  This goal has since evolved.
Current plans for observing star clusters in the SDSS filter system
now include: (1) their potential use in calibrating the SDSS extension
programs, and (2) creating a large uniform imaging survey of clusters
similar to the WIYN Open Cluster Survey (WOCS)
(see Mathieu 2000).

In the first of our cluster papers (Rider et al.\ 2004), we present results
for NGC~2548.  We chose this particular cluster because we had
observations of it using three telescopes: the CTIO Curtis-Schmidt,
the PT, and, most importantly, the USNO-1~m, the telescope with
defines the $u'g'r'i'z'$ photometric system (Smith et al.\ 2002).  
A second cluster paper (Moore et al.\ 2006) has been submitted and
addresses NGC~6134 and Hogg-19.  Further papers are in preparation.
In all, we have data in hand for $\approx$100 open clusters.

These data may be used to verify the recent age and metallicity models
of Girardi et al.\ (2004) and the prior work of Lenz et al.\ (1998), and to
verify and expand upon the $u'g'r'i'z'$ to $UBVR_cI_c$ transformations
presented in the standard star paper (Smith et al.\ 2002).
Rodgers et al.\ (2006) is taking another look at these filter system
transformations for main sequence stars in an effort to sort out
luminosity class effects.

\subsection{SDSS-II}

The SDSS officially ended on 2005 June 30.  The SDSS-II is a 3-year
extension of the SDSS beyond this date, and is composed of three
parts:
\begin{enumerate}
\item The Legacy Survey, which will fill in a gap in the SDSS imaging
	and spectroscopic coverage of the Northern Galactic Cap due to
	worse-than-expected weather conditions during the five years
	of the original SDSS.
\item The Sloan Extension for Galactic Understanding and Exploration, 
	or SEGUE, which will image about 3500 sq deg of sky --- mostly
	at Galactic latitudes of $|b^{II}| < 30^{\circ}$ --- and measure
	spectra to determine abundances and radial velocities of a 
	quarter of a million stars, in order to study the structure
	and evolution of the Milky Way (Newberg et al.\ 2004).
\item The SDSS-II Supernova Survey, which will image 250 sq deg along
	the Celestial Equator every two nights during the Autumns of
	2005-2007, with the aim of obtaining well-sampled,
	well-calibrated, multi-band lightcurves for 200 Type Ia
	supernovae in the redshift ``desert'' $z = 0.1-0.35$
	(Frieman et al.\ 2004; Sako et al.\ 2005).
\end{enumerate}

The PT and {\tt MTPIPE} are playing an important role in all three of
these components of the SDSS-II.  First of all, since it is basically
a continuation of the original SDSS within the original survey
borders, the Legacy component of SDSS-II continues to benefit from
the use of {\tt MTPIPE} reductions of PT secondary patches as part of
the general photometric calibration strategy of the 2.5m imaging
camera data.

Likewise, the 2.5m imaging scans for SEGUE also uses {\tt MTPIPE}
reductions of PT secondary patches located within the SEGUE survey
area (see Fig.~\ref{secpatchlocationsRaDec}).  But the PT and other
small telescopes (like the USNO-1.0m) have an even bigger role to
play in SEGUE.  One of the major thrusts of SEGUE is the SEGUE Open
Cluster Survey.  This is a wide-field imaging program with
limited spectroscopic follow-up within SDSS-II.  While the imaging
program will not go as faint as the WOCS
(see Mathieu 2000), we will observe almost an order of magnitude
more clusters than the WOCS.  Directed pointings for specific calibration
targets and for the brighter stars of the clusters will be undertaken
with the smaller telescopes, and those data will be reduced with {\tt
MTPIPE}.

Finally, the PT and {\tt MTPIPE} have a role to play in the Supernova
component of the SDSS-II.  In this case, the role has less to do with
photometric calibrations --- since the primary Supernova strategy is
to use differential photometry of repeat 2.5m imaging scans
against previously calibrated areas of the sky scanned by the original
SDSS survey --- but more of a role in followup observations.  Here, the
PT will be able to follow the light curves of the brighter candidate
Type Ia supernovae detected by the main SDSS-II Supernova Survey during
those times when 2.5m telescope time is allotted either to the Legacy
survey or to SEGUE, and {\tt MTPIPE} will be able to process these PT
observations.

\section{The Future}
\label{sec:conclusions}

{\tt MTPIPE} is a mature and stable data processing pipeline.  The
last major revision --- from {\tt v7.4} to {\tt v8.0} occurred over
four years ago (on 2002 April 22).  As noted in \S~\ref{sec:packages},
the current version is {\tt v8.3}, and the changes since {\tt v8.0}
have been mostly cosmetic and have generally dealt with the smooth
running of the PT Factory (\S~\ref{sec:ptfactory}).  This being the
case, what is the future of {\tt MTPIPE}?

There are, of course, many relatively small changes to {\tt MTPIPE}
that could yield incremental improvements in the final output and to
the photometric calibrations of the SDSS.  These include updates to
the de-fringing code in {\tt mtFrames} and continued improvements to
the automated quality assessment of secondary patches for the PT
Factory.  Also, improving the compatibility of the output of {\tt
MTPIPE} with other astronomical data packages would be a plus.  Along
these lines, adding World Coordinate System (WCS;
Calabretta \& Greisen 2002; Greisen \& Calibretta 2002) keywords to the FITS headers of
{\tt MTPIPE} corrected images is high on the list of priorities.
(Currently, there is a standalone Tcl/{\tt ASTROTOOLS} script to do
this, but it has yet to be fully incorporated into {\tt MTPIPE}.)

A larger task that may be ahead for {\tt MTPIPE} is how to deal with
crowded field photometry.  {\tt mtFrames} is an aperture photometry
code, but many of the science projects that currently use or plan to
use {\tt MTPIPE} involve crowded fields --- in particular, star
clusters (\S~\ref{sec:otherprojects}).  Furthermore, the photometric
calibration SDSS-II/SEGUE scans may require the placement of SEGUE
secondary patches at very low Galactic latitude.  With the task of
crowded field photometry in mind, then, we are investigating the
possibility of adding a toggle switch to {\tt mtFrames} which would
allow one to choose to run a PSF-fitting photometry package like
DAOPHOT (Stetson 1987) or DoPHOT (Schechter et al.\ 1993) on corrected
frames output by {\tt mtFrames}; a new script, which would be run
after {\tt mtFrames}, would then package the resulting DAOPHOT/DoPHOT
output into an {\tt excal}-/{\tt kali}-ingestible format.

\acknowledgements

%%SDSS acknowledgment blurb as of 30 June 2005:
%Funding for the Sloan Digital Sky Survey (SDSS) has been provided by
%the Alfred P. Sloan Foundation, the Participating Institutions, the
%National Aeronautics and Space Administration, the National Science
%Foundation, the U.S. Department of Energy, the Japanese
%Monbukagakusho, and the Max Planck Society. The SDSS Web site is
%{\tt http://www.sdss.org/}.
%
%The SDSS is managed by the Astrophysical Research Consortium (ARC) for
%the Participating Institutions. The Participating Institutions are The
%University of Chicago, Fermilab, the Institute for Advanced Study, the
%Japan Participation Group, The Johns Hopkins University, the Korean
%Scientist Group, Los Alamos National Laboratory, the
%Max-Planck-Institute for Astronomy (MPIA), the Max-Planck-Institute
%for Astrophysics (MPA), New Mexico State University, University of
%Pittsburgh, University of Portsmouth, Princeton University, the United
%States Naval Observatory, and the University of Washington.

%SDSS acknowledgment blurb as of 26 June 2006:
Funding for the SDSS and SDSS-II has been provided by the Alfred P.
Sloan Foundation, the Participating Institutions, the National Science
Foundation, the U.S.  Department of Energy, the National Aeronautics
and Space Administration, the Japanese Monbukagakusho, the Max Planck
Society, and the Higher Education Funding Council for England.  The
SDSS Web Site is {\tt http://www.sdss.org/}.

The SDSS is managed by the Astrophysical Research Consortium for the
Participating Institutions. The Participating Institutions are the
American Museum of Natural History, Astrophysical Institute Potsdam,
University of Basel, Cambridge University, Case Western Reserve
University, University of Chicago, Drexel University, Fermilab, the
Institute for Advanced Study, the Japan Participation Group, Johns
Hopkins University, the Joint Institute for Nuclear Astrophysics, the
Kavli Institute for Particle Astrophysics and Cosmology, the Korean
Scientist Group, the Chinese Academy of Sciences (LAMOST), Los Alamos
National Laboratory, the Max-Planck-Institute for Astronomy (MPIA),
the Max-Planck-Institute for Astrophysics (MPA), New Mexico State
University, Ohio State University, University of Pittsburgh,
University of Portsmouth, Princeton University, the United States
Naval Observatory, and the University of Washington.

%The GSC acknowledgment blurb:
The Guide Star Catalog-I was produced at the Space Telescope Science
Institute under U.S. Government grant. These data are based on
photographic data obtained using the Oschin Schmidt Telescope on
Palomar Mountain and the UK Schmidt Telescope.

The Guide Star Catalogue-II is a joint project of the Space Telescope
Science Institute and the Osservatorio Astronomico di Torino. Space
Telescope Science Institute is operated by the Association of
Universities for Research in Astronomy, for the National Aeronautics
and Space Administration under contract NAS5-26555. The participation
of the Osservatorio Astronomico di Torino is supported by the Italian
Council for Research in Astronomy. Additional support is provided by
European Southern Observatory, Space Telescope European Coordinating
Facility, the International GEMINI project and the European Space
Agency Astrophysics Division.

We also thank Arlo Landolt, Nick Suntzeff, Arne Henden, and Leo Girardi 
for insightful discussions during the development of the standards star
project.  Most of these discussions are manifested in improvements to
the software.
 
Finally, we thank the anonymous referee for reviewing this paper and
providing us with useful comments.

%\clearpage

\end{document}